\documentclass[journal]{IEEEtran}
\usepackage{cite}
\usepackage{amsmath,amssymb,amsfonts}
\usepackage{graphicx}
\usepackage{subfigure}
\usepackage{textcomp}
\usepackage[table,xcdraw]{xcolor}
\usepackage{float}
\usepackage{algorithm}
\usepackage{algpseudocode}
\usepackage{url}
\usepackage{multirow}
\usepackage{slashbox}
\usepackage{diagbox} 
\usepackage{threeparttable}
\usepackage{booktabs}
\ifCLASSINFOpdf
\else
\fi

\begin{document}
\title{DeepKeyGen: A Deep Learning-based Stream Cipher Generator for Medical Image Encryption and Decryption}
\author{Yi Ding, ~\IEEEmembership{Member,~IEEE,}, Fuyuan Tan, Zhen Qin, Mingsheng Cao, Kim-Kwang Raymond Choo~\IEEEmembership{Senior Member,~IEEE,} 
	and Zhiguang Qin, ~\IEEEmembership{Member,~IEEE}
	
\thanks {Corresponding author: Zhen Qin (qinzhen@uestc.edu.cn)}
\IEEEcompsocitemizethanks{
\IEEEcompsocthanksitem Yi Ding is with the Network and Data Security Key Laboratory of Sichuan Province, University of Electronic Science and Technology of China, Chengdu, Sichuan, 610054 China; he is also with Institute of Electronic and Information Engineering of UESTC in Guangdong, Guangdong, 523808, China (e-mail: yi.ding@uestc.edu.cn).
\IEEEcompsocthanksitem Fuyuan Tan, Zhen Qin, Mingsheng Cao and Zhiguang Qin are with the Network and Data Security Key Laboratory of Sichuan Province, University of Electronic Science and Technology of China, Chengdu, Sichuan, 610054 China (e-mail: tanfuyuan@std.uestc.edu.cn; qinzhen@uestc.edu.cn; cms@uestc.edu.cn; qinzg@uestc.edu.cn).
\IEEEcompsocthanksitem Kim-Kwang Raymond Choo is with the Department of Information Systems and Cyber Security, University of Texas at San Antonio, San Antonio, TX 78249-0631, USA (e-mail: raymond.choo@fulbrightmail.org).}
}
\markboth{}%
{}\maketitle

\begin{abstract}
The need for medical image encryption is increasingly pronounced, for example to safeguard the privacy of the patients' medical imaging data. In this paper, a novel deep learning-based key generation network (DeepKeyGen) is proposed as a stream cipher generator to generate the private key, which can then be used for encrypting and decrypting of medical images. In DeepKeyGen, the generative adversarial network (GAN) is adopted as the learning network to generate the private key. Furthermore, the transformation domain (that represents the ``style'' of the private key to be generated) is designed to guide the learning network to realize the private key generation process. The goal of DeepKeyGen is to learn the mapping relationship of how to transfer the initial image to the private key. We evaluate DeepKeyGen using three datasets, namely: the Montgomery County chest X-ray dataset, the Ultrasonic Brachial Plexus dataset, and the BraTS18 dataset. The evaluation findings and security analysis show that the proposed key generation network can achieve a high-level security in generating the private key.
\end{abstract}


%
\IEEEpeerreviewmaketitle

\section{Introduction}
\IEEEPARstart{A}{s} medical imaging becomes increasingly commonplace, so does the use of medical images to inform diagnosing and treatment plans, etc. For example, images from brain magnetic resonance imaging (MRI) and computed tomography (CT) of chest can be used to facilitate brain tumor detection for lung diagnosis. However, these medical images contain sensitive and private information about the patients, and their leakage can have potential privacy implications for the patients and legal ramifications for the hospitals. Hence, there has been efforts devote to designing security solutions (e.g. cryptographic primitives) to secure these medical images and protect the patients' privacy.   

Stream and block ciphers are two examples of popular cipher systems used in medical image encryption algorithm.

\begin{figure}[H]
	\centerline{\includegraphics[width=\columnwidth]{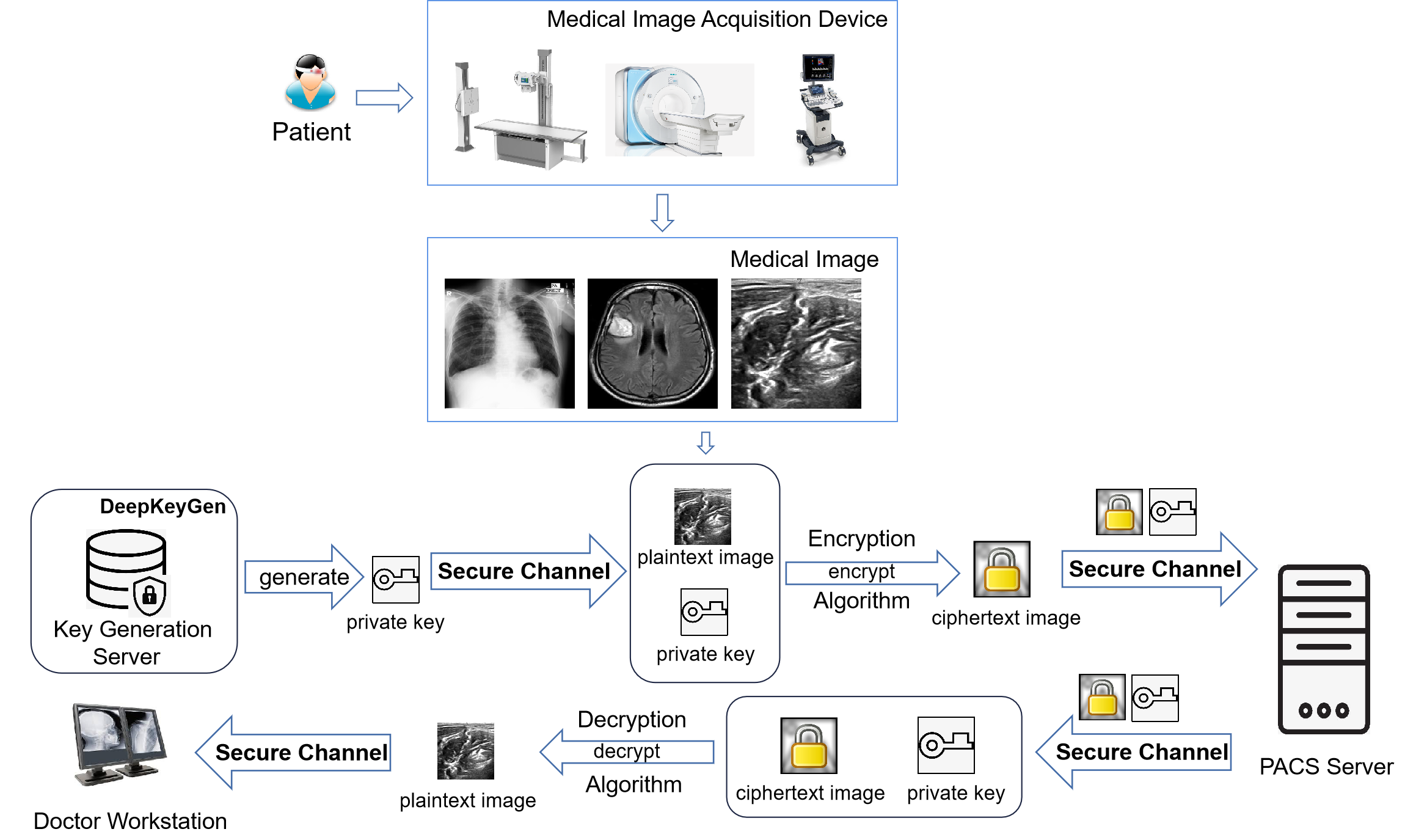}}
	\caption{An example application scenario of proposed DeepKeyGen.}
	\label{fig-app}
\end{figure}

\noindent Compared to block ciphers (e.g., Data Encryption Standard (DES), International Data Encryption Algorithm (IDEA), and Advanced Encryption Standard (AES)), stream ciphers generally have a high-level security, are faster in terms of encryption and decryption speed, have small error expansion, achieve better synchronization, and incur lower implementation cost \cite{a1,a2,a3}. One challenge, however, is how to design a security stream cipher generator to facilitate the process of generating the randomized and unpredictable sequence. Common stream cipher generators include linear feedback shift register \cite{e1}, nonlinear feedback shift register \cite{e2}, finite automation \cite{e3}, linear congruence generators \cite{e4} and chaotic systems \cite{e5}. In most existing approaches involving the use of private key generators, the generators are manually designed (e.g., using mathematical formulas) to generate the private key in a specific ``style'' in order to achieve a certain security level. Then, the generators are independently realized by repeated attempts until the generated private key achieves or approaches the expected ``style''. However, for each attempt, the implementation process usually requires one to adjust the calculation model manually and it is challenging for the generator  to achieve the best expected performance as designed. Moreover, the designing and implementing processes are time-consuming work and costly (e.g., human experts involvement). Therefore, instead of manually designing and implementing the key generator, this paper mainly focuses on how to realize the private key generator in a learning manner. To be more specific, if we know the desired ``style'' of the private key, a deep learning network can be trained to learn how to generate the expected private key from the ``seed''. In other words, we posit the potential of using deep learning to automatically and securely generate the private key.

Deep learning has been successfully utilized to a wide variety of computer vision tasks \cite{d1,d2,d3,d4,d5,d6}, and generative adversarial network (GAN) \cite{b8} is one of the most popular deep learning algorithms. GAN consists of the generator and the discriminator, where the generator is responsible for generating samples and the discriminator learns to distinguish between generated sample and sample from real data. The generator and the discriminator compete with each other to make the generated data as realistic as possible. GAN-based methods are efficiently in image-to-image translation, where images are transferred from one domain to another. This can also be considered as a type of computer vision tasks, whose goal is to learn the mapping relationship between two image domains.

Therefore, we design a deep learning based key generation network (DeepKeyGen) by fusing both stream cipher generator and image-to-image translation. This concept is based on the following two significant observations. 
\begin{enumerate}
	\item If we know the desired ``style'' of the private key, then this particular style can be set as the transformation domain. The image-to-image translation is able to build the mapping relationship between two image domains and transfer the images from the source domain to the transformation domain (i.e., learn how to achieve the required ``style'' of the private key). In this way, we can regard the image-to-image translation as a process of generating the private key, and the output image is actually a private key that can be used to encrypt the medical images.
	\item Deep learning model has a large number of parameters, complex network structure and random training process. Hence, it can be an alternative method for stream cipher generator.
\end{enumerate}

In DeepGenKey, the GAN is employed as the learning network. There are two domains in the generating process, namely: source and transformation domains. The source domain can be any images that have the same distribution. The initial image is from the source domain, which is adopted as the ``seed'' to generate the private key. The transformation domain represents the ``style'' of the private key to be generated, such as chaotic private key with a certain level of security. During training, the generator transfers the images from the source domain to the transformation domain, and the output of generator is regarded as the private key. The discriminator is used to distinguish between the key generated by the generator and the data from the transformation domain. Due to the randomness of the training process for deep learning, the generated private keys differ even under the same training conditions. In other words, the proposed DeepKeyGen can be considered as a one-time pad. Moreover, DeepGenKey utilizes unlabeled and unpaired images to train the learning network; thus, overcoming data availability issue in training the GAN. 

Fig.\ref{fig-app} is an example of how DeepKeyGen can be applied to encrypt/decrypt medical images in a healthcare setting. For example, upon receiving the medical image from patients, the private key is generated by the key generation server. Then, we can use the encryption algorithm to encrypt the unencrypted image (medical image) with the generated private key, and thus obtain the corresponding ciphertext (encrypted medical image). Subsequently, the ciphertext and the generated private key are transferred via a secure channel and stored in the Picture Archiving and Communication Systems (PACS) server. When a medical doctor (or another authorized healthcare worker) wishes to view the medical image, the encrypted image and the corresponding private key are first retrieved from the PACS server. The decryption algorithm is then used to decrypt the encrypted image with the corresponding private key, in order to obtain the unencrypted (original) image. The unencrypted image is subsequently sent via a secure channel to the medical doctor's workstation and the medical doctor can view the original medical image. We assume that the systems run in an intranet environment and thus, the transmissions are carried out via the secure channel.

We will now summarize the key contributions of this work as follows:
\begin{enumerate}
	\item Design a novel deep learning based private key generator, DeepKeyGen, to realize the key generation process using deep learning in image-to-image transformation. The private key generated by the proposed method has a large key space and high randomness, and is also sensitive to changes. To the best of our knowledge, the proposed method is one of the earliest works to utilize deep learning to realize private key generator in a learning approach. Moreover, instead of manually designing and implementing the key generator, this work presents a new research direction, and more specifically using learning to automatically realize private key generator with high security level.
	\item Carry out extensive experiments using Montgomery County's chest X-ray dataset \cite{ds1}, the Ultrasonic Brachial Plexus dataset \cite{ds2}, and the BraTS18 dataset \cite{ds3,ds4} to evaluate the utility of DeepKeyGen. These three datasets represent the most commonly used modalities (X-Ray, Ultrasound and MRI) in clinical practice. Findings from our evaluations demonstrate that the generated private key has the high-level security and randomness. Moreover, the plaintext medical images with multi-modality can be encrypted efficiently using the generated private key, and the generality of the proposed DeepKeyGen is also been demonstrated. In addition, we show that DeepKeyGen is resilient to various known attacks, even if the attacker knows the entire private key generation process.
\end{enumerate} 

In the next section, we will introduce the related literature, prior to presenting our proposed DeepKeyGen in the third section. We then describe our security and performance evaluations in the next two sections, before concluding this paper in the last section.

\section{Related Work}
\subsection{Existing Key Generation Algorithms}

A number of key generation techniques have been proposed in the literature. For example, Moosavi \emph{et al.} \cite{b1} proposed a secure low-latency key generation method based on ECG features. This method uses Fibonacci linear feedback shift registers and AES algorithms as pseudo-random number generators to achieve a high security level. Liu \emph{et al.} \cite{b2} proposed an authentication and dynamic encryption method based on state estimation, which selects the metrics of the power system to generate the key. As the value of power measurement used to generate the key is constantly changing and unpredictable, the key is challenging to guess. Kalsi \emph{et al.} \cite{b3} demonstrated the possibility of hiding data based on the combination of DNA sequences and deep learning, where their proposed key generation approach uses the natural selection theory, Needleman-Wunsch algorithm, and genetic algorithm.

There have also been interest in exploring the utility of non-linear systems, such as chaotic system (known to have properties such as sensitivity, unpredictability, pseudo-randomness, ergodicity and certainty \cite{f1,f2,f3}), in key generation. For example, Garcia-Bosque \emph{et al.} \cite{b4} adopted a novel sensor-based ``seed'' generator to generate a stream cipher key, by combining a hybrid tilt tent graph and a linear feedback shift register algorithm. The proposed method reportedly achieves good security and efficiency. Zahmoul \emph{et al.} \cite{b5} proposed a new chaotic graph based on Beta functions to generate chaotic sequences, which are different from commonly used ones. The generated chaotic sequences are mainly used to shuffle the image's pixel positions and to conceal the relationship between the encrypted original images. Arab \emph{et al.} \cite{b6} introduced a novel image encryption algorithm, by fusing chaotic sequence and an improved AES algorithm. The private key is generated from an Arnold chaotic sequence, and the plaintext image is then encrypted using the improved AES algorithm and the round keys generated by the chaotic system. It was shown that the key space is sufficiently large to mitigate brute force attacks. Lambić \emph{et al.} \cite{b7} explained how one can generate pseudo-random numbers using discrete-space chaotic mapping, and showed that the generated private key achieves key randomness and has an infinite key space. The proposed approach can also reportedly generate the same number of different pseudo-random sequences as other secure discrete space chaos methods, while incurring minimal memory space.

These existing key generation algorithms require the generators to be manually designed, and the generator realizing process is then repeated several times and the underlying mathematical formulas constantly refined to achieve or approach the desired style. This is clearly time- and resource-expensive. 

\subsection{Deep Learning-based Image-to-Image Transformation}
Deep learning is a recent research trend in a broad range of applications (including image processing tasks, such as image classification \cite{d1,d2}, object detection \cite{d3,d4} and image segmentation \cite{d5,d6}), since its multi-layer network structure can be used to effectively express complex functions. Generative adversarial network (GAN) \cite{b8} is a branch of deep learning, and a GAN algorithm generally consists of the generator and the discriminator. The generator is responsible for capturing the distribution of the sample data, and the discriminator is responsible for determining whether the input is real data or a generated sample. Since the seminal work of Goodfellow \emph{et al.} \cite{b8} in 2004, many GAN-based methods have been designed for different applications \cite{g1,g2,g3}. Lsola \emph{et al.} \cite{b9}, for example, proposed a supervised image-to-image translation structure Pix2Pix based on CGAN. The Pix2Pix employs the real image as an additional input information to the generator. Moreover, it adds the $L_1$ loss as the penalty term for the generator, in order to improve the generator to generate more realistic results.

Inspired by the concept of cycle consistency, Zhu \emph{et al.} \cite{b10} proposed the CycleGAN model to solve the problem of special training dataset for style transfer tasks. It can be trained by adopting the unpaired and unlabeled data, so as to facilitate the application of GAN in image-to-image transformation. Kim \emph{et al.} \cite{b11} proposed DiscoGAN to ensure that certain features of the image are preserved when the image is transferred to another domain. Inspired by the original dual learning method of natural language processing, Yi \emph{et al.} \cite{b12} proposed the DualGAN model, which can be used to translate images between two domains with different characteristics by using unlabeled and unpaired data.

Although deep learning methods have been widely adopted in the area of image-to-image transformation and other application domain, only a few researchers have attempted to employ deep learning algorithms in private key generation. As discussed earlier, we will demonstrate that deep learning can be used to automatically realize private key generation that results in a private key with a high security level. 

\section{Our Proposal}

\subsection{Encryption and Decryption Architecture}

\begin{figure}[h]
	\centerline{\includegraphics[width=\columnwidth]{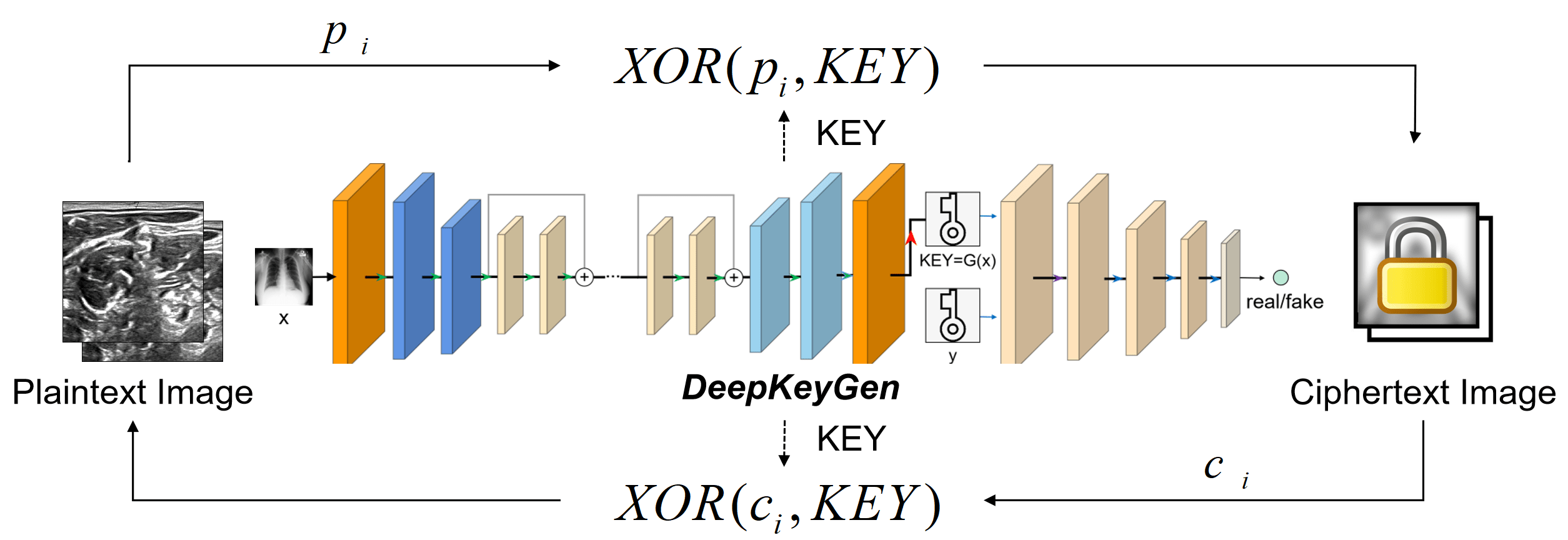}}
	\caption{DeepKeyGen-based encryption and decryption system.}
	\label{fig1}
\end{figure}

Fig.\ref{fig1} presents an overview of the encryption and decryption system based on DeepKeyGen. This encryption and decryption system combines the stream cipher generated by DeepKeyGen with a XOR algorithm. During encryption, the unencrypted image (hereafter referred to as plaintext,andor denoted as $p_i$) is encrypted using the generated private key and the XOR algorithm. As a result, we obtain the encrypted image (hereafter referred to as ciphertext, and denoted as $c_i$). Decryption is the reverse of encryption. The variable y represents the transformation domain, which is used as the ``ground truth'' in the discriminator network.

\subsection{DeepKeyGen}

\begin{figure*}[]
	\centerline{\includegraphics[width=1\textwidth]{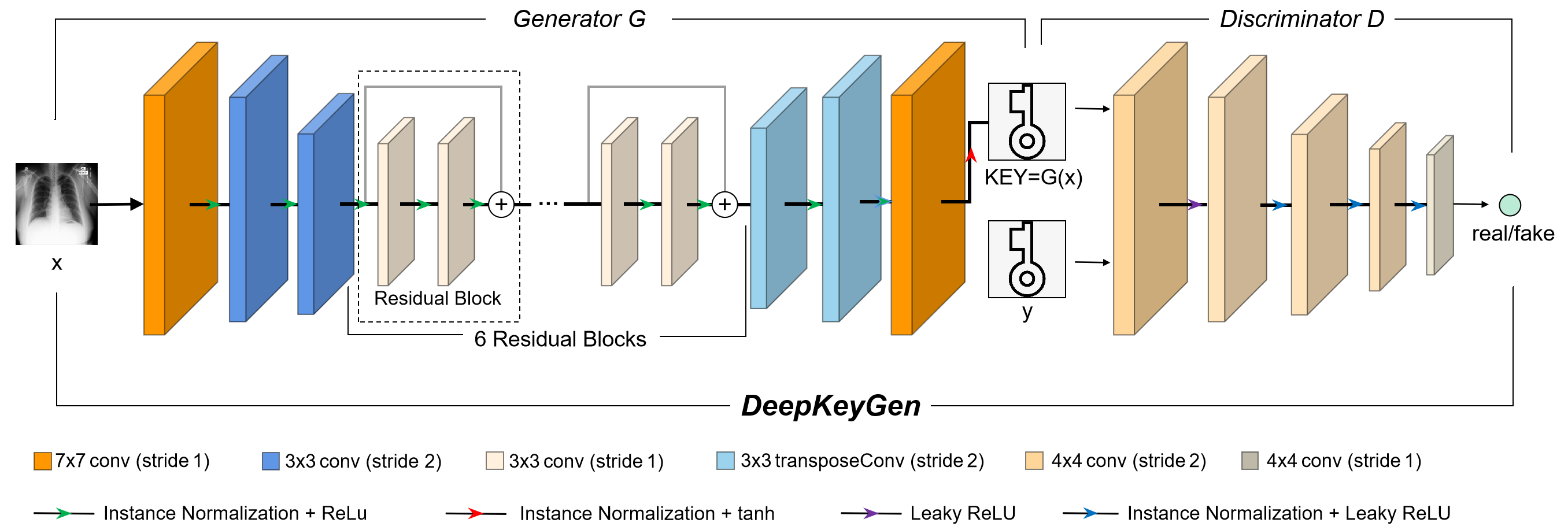}}
	\caption{Proposed DeepKeyGen architecture. The proposed DeepKeyGen is a deep learning network, which consists of a generator and a discriminator to generate the private key. The input of the generator network is the ``seed'' images obtained from the source domain. The y represents the target image obtained from the transformation domain, which indicates the ``style'' of private key to be generated. The discriminator network is used to determine whether the generated key belongs to the transformation domain.}
	\label{fig2}
\end{figure*}

As shown in Fig.\ref{fig2}, DeepKeyGen comprises a generator G and a discriminator D. The generator network G is used to generate private key from the initial image, which is in the form of an image. The discriminator network D is responsible to distinguish between the private key generated by the generator and the real data from transformation domain. In DeepKeyGen, both source and transformation domains facilitate key generation. Specifically, the source domain contains the initial images, which are adopted as the ``seed'' to generate the private key. The transformation domain contains the target image, where the initial image wants to transfer. Furthermore, the transformation domain represents the ``style'' of the private key to be generated, such as a chaotic private key with high security level. In addition, the loss function is usually adopted to train the deep learning network and the total loss function $L$ used to train DeepKeyGen is described as follows:

\begin{equation}
L = {L_G} + {L_D}
\end{equation}
In the above equation, ${L_G}$ and ${L_D}$ respectively denote the loss functions of generator network G and discriminator network D.

\subsubsection{Generator Network G}

The generator network G is used to transfer the initial image from source domain onto the transformation domain. The output of the generator network G is the private key which holds the same security characteristics as the encrypted performance in the transformation domain. The generator network G is composed of three downsample layers, six residual blocks, two transposed convolutional layers and an ordinary convolutional layer. The initial image is firstly processed with three downsample layers. These layers are aimed at extracting the features from the initial image. Then there are six residual blocks with the same structure \cite{c1} to construct the content and diverse features. Next, the feature vectors pass through two transposed convolutional layers where the goal is to revert them to the low-level dimension. Finally, the low-level features are converted into an image by a normal convolution. And the output image is regarded as the generated private key for the DeepKeyGen. For each convolutional layer, the instance normalization is employed to implement the normalization operation on a single image so as to improve the quality of the generated image, while accelerating model convergence and preventing gradient explosion. The loss function of the generator network G is:		

\begin{equation}
{L_G} = \mathop {{\rm{min}}}\limits_G ({{\rm E}_{x  \sim pdata(x)}}\log (1 - D(G(x)))
\end{equation}
In the above equation, $G$ denotes the generator network, $D$ represents the discriminator network, and $x$ represents the initial image (also known as the ``seed''). The loss function ${L_G}$ can be understood as the key generated by the generator ``misleading'' the discriminator to the maximum extent. It means that the generated key is getting close to the transformation domain, and the discriminator believes that the generated key comes from the transformation domain.

\subsubsection{Discriminator Network D}

The discriminator network D is used to determine whether the generated image belongs to the transformation domain. The discriminator network D is composed of five convolutional layers. The input image is sequentially handled by four convolutional layers to extract useful features. Then, the features are processed by the last layer, and output a one-dimensional result. The latter represents either real or fake, and facilitate discrimination. Besides the first and last layers, instance normalization is also used to implement normalization. The loss function of the discriminator is  as follows:

\begin{equation}
\begin{split}
{L_D} = \mathop {{\rm{max}}}\limits_D ({{\rm E}_{{\rm{y}} \sim pdata({\rm{y}})}}\log (D(y)) + \\
{{\rm E}_{{\rm{x}} \sim pdata({\rm{x}})}}\log (1 - D(G(x))))
\end{split}
\end{equation}
In the above equation,  $G$ represents the generator network, $D$ represents the discriminator network, $x$ represents the initial image (``seed'') from the source domain, and $y$ represents the data from the transformation domain. The loss function ${L_D}$ can be understood as the maximization of the classification accuracy of the discriminator.

In  GAN, both ${L_G}$ and ${L_D}$ form the adversarial system. When the two networks reach a balance, the accuracy of the discriminator network D should be close to 50\%. In other words, the generated key is similar to that from the transformation domain, and consequently the discriminator network D is not able to distinguish between the two.

\subsection{Image Private Key}

The private key generated by DeepKeyGen is a type of stream ciphers, and is also an image. For each image, it consists of a sequence of pixels. These pixels not only contain the information of pixel value, but also hold the spatial information. Therefore, the private key can be defined as a composite of image pixels, which is described as follows: 

\begin{equation}
KEY_{definition} = [{V_1},{V_2},...{V_i},...{V_n}]
\end{equation}
In the above equation, $V_i$ represents one pixel in the image. It also represents one value in the key sequence. Each $V_i$ is composed of a quadruples and is defined as follows:

\begin{equation}
{V_i} = \left[ {{p_i},x,y,c} \right]
\end{equation}
In the above equation, ${p_i}$ represents the pixel value, $x$ is the horizontal coordinate value, $y$ is the vertical coordinate value, and $c$ is the the RGB color channel information. The value ranges of ${p_i}$, $x$ and $y$ are from 0 to 255, and the value range of $c$ is from 0 to 2. The private key differs from that of existing stream ciphers, which is actually a four-dimensional key. The key values (which contain the pixel value information and the three-dimensional space position information) ensure the private key is complex and the key space is sufficiently large; thus, significantly enhancing the key's security level.

\subsection{Key Generation Process}

The key generation process is actually a network training process for learning the mapping relationship from the source domain to the transformation domain. Before the training, the parameters of each convolutional layer in DeepKeyGen are randomly initialized as follows:

\begin{equation}
{W_n} = random\left[ {{w_{n,1}},{w_{n,2}},...,{w_{n,i}}} \right]
\end{equation}
In the above equation, ${{w_{n,i}}}$ represents the $i$th parameter of the nth convolution layer of the DeepKeyGen. All parameters of DeepKeyGen $W$ are composed of the parameters from all convolutional layers, which are defined as follows:

\begin{equation}
W = consist\left[ {{W_1},{W_2},...,{W_n}} \right]
\end{equation}

The proposed DeepKeyGen comprises a generator and a discriminator, where the generator is used to generate the private key and can be expressed as follows:

\begin{equation}
KEY = G(W;x)
\end{equation}
In the above equation, $G()$ represents the convolutional neural network of the generator, $W$ indicates all parameters in the network, and $x$ denotes the initial image. 

\begin{figure}[ht]	
	\centering
	\subfigure{
		\includegraphics[width=\columnwidth]{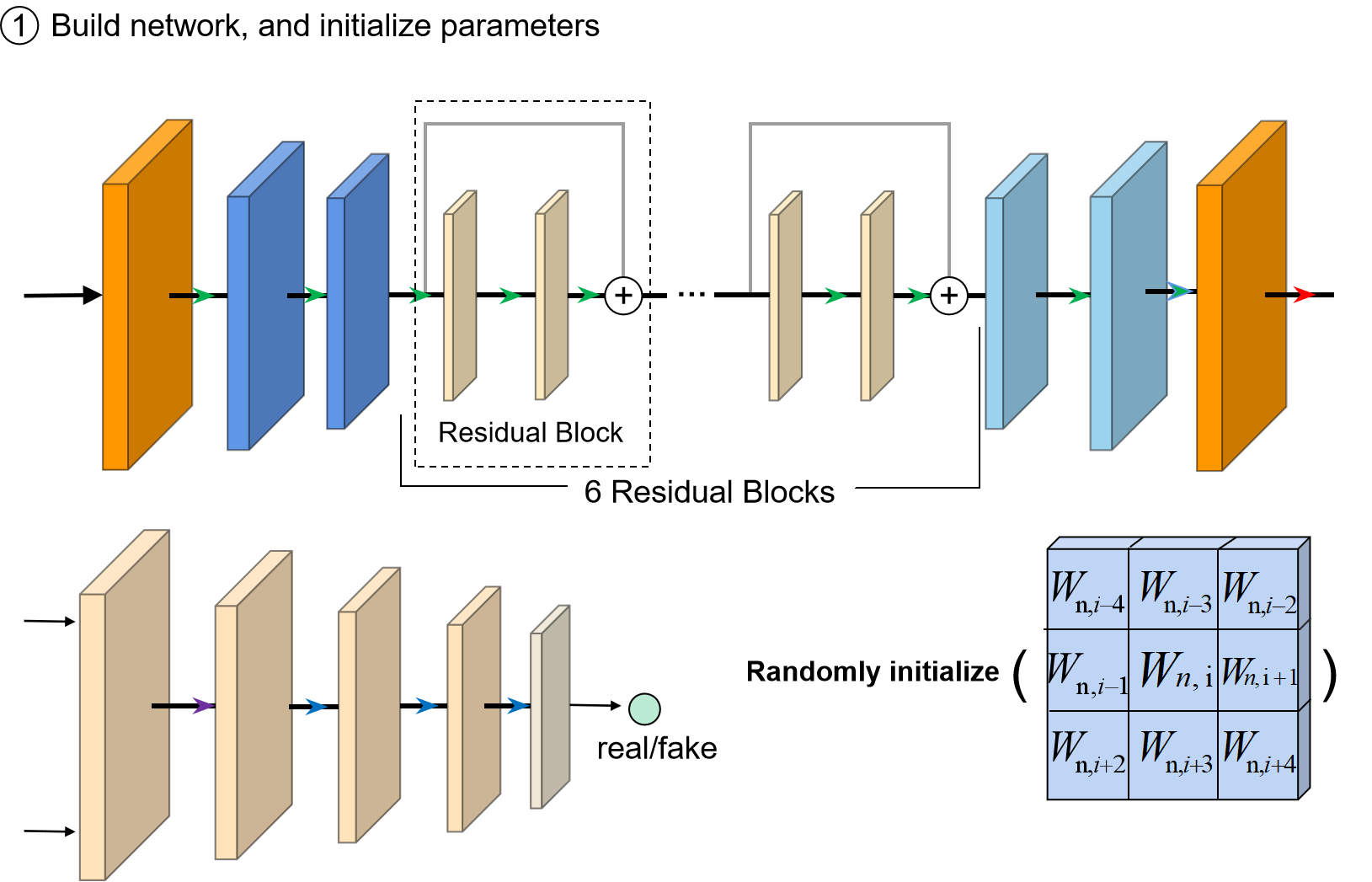} 
	}
	\subfigure{
		\includegraphics[width=\columnwidth]{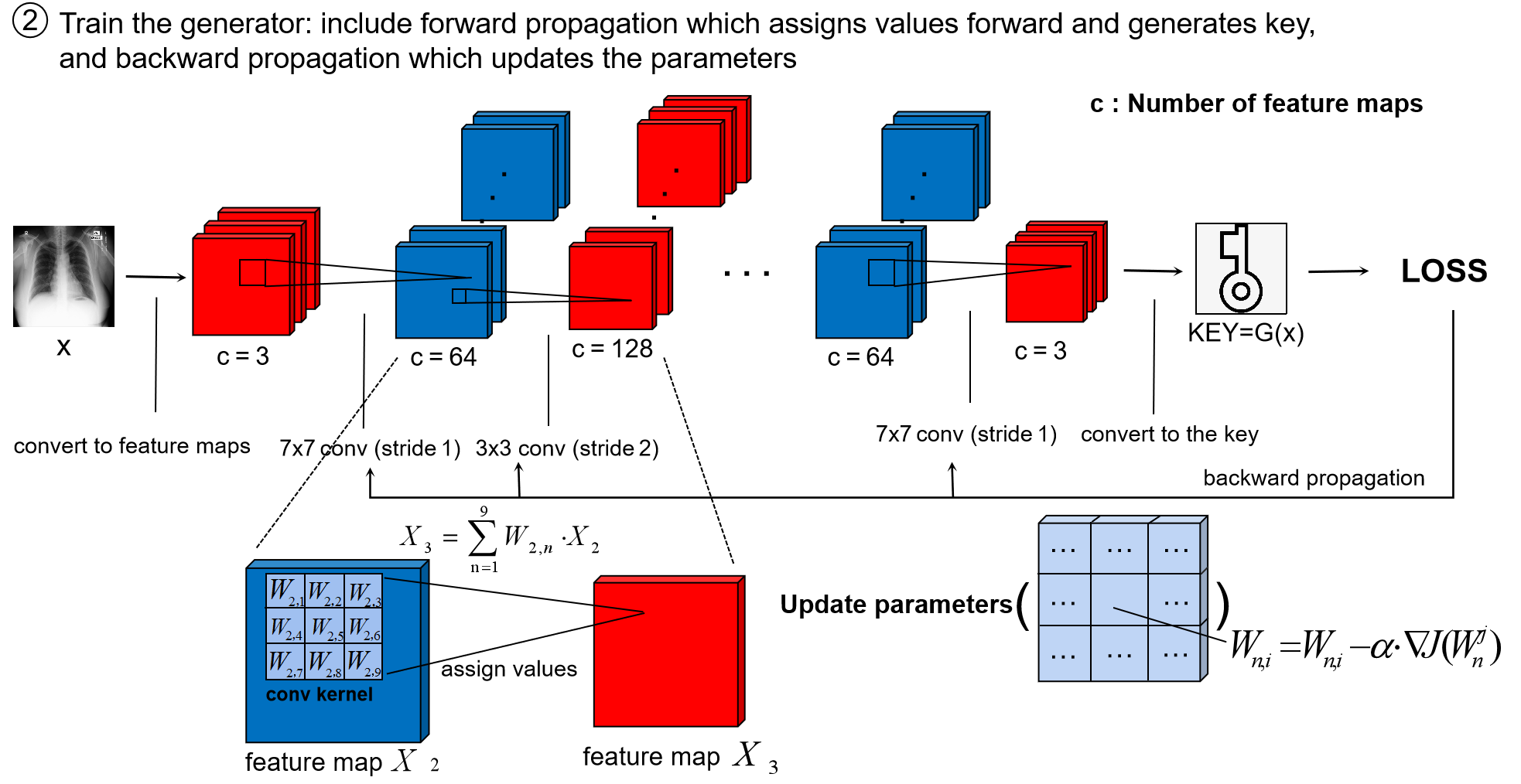} 
	}
	\subfigure{
		\includegraphics[width=\columnwidth]{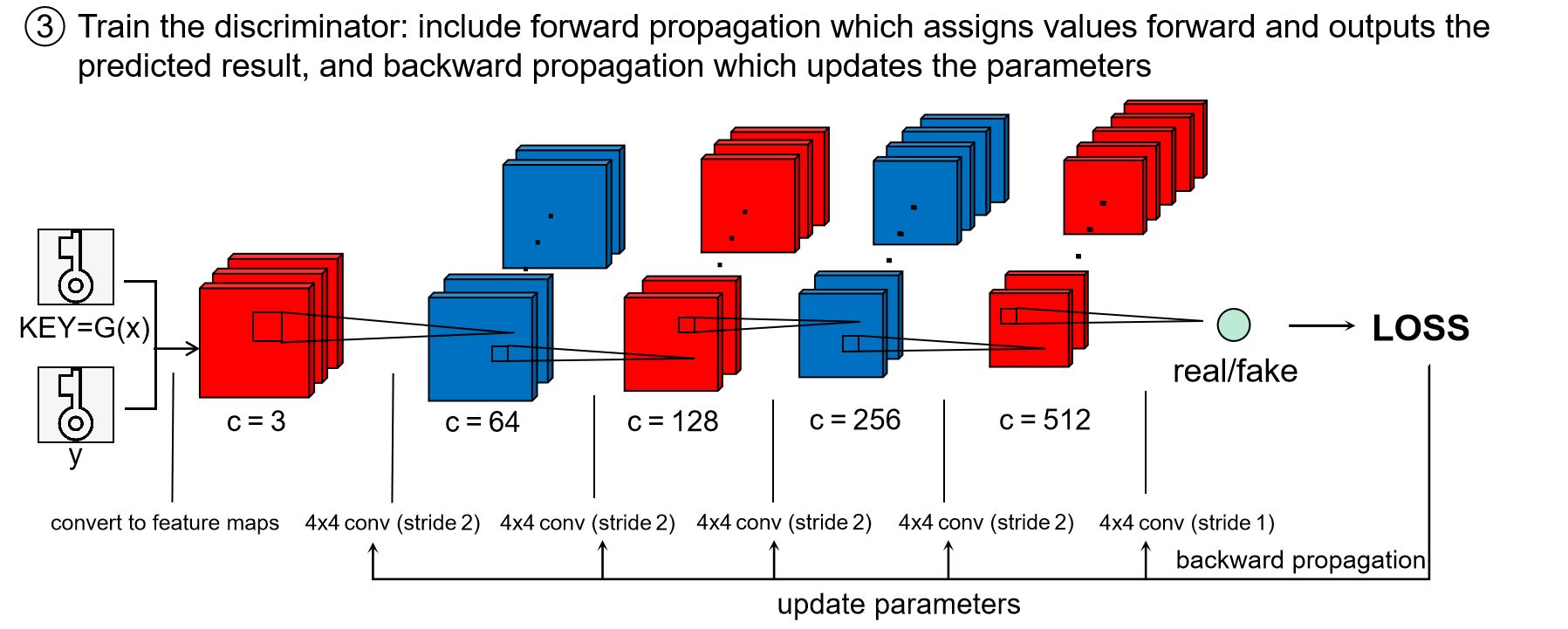} 
	}
	\caption{Proposed DeepKeyGen's four-step key generation process. Step 1: Constructs the DeepKeyGen network with a generator network (top) and a discriminator network (bottom left). In addition, it randomly initializes the network's parameters. Step 2: Trains the generator with the forward propagation, and generates the original private key based on the ``seed'' image. Step 3: Trains the discriminator to determine whether the generated key belongs to the transformation domain. We remark that both steps 2 and 3 are simultaneously processed. Step 4: The backward propagation is carried out to update the parameters of both generator and discriminator networks, until it achieves the best generation performance.}
	\label{fig5}
\end{figure}

At the beginning of training, the input image is converted into feature vectors through the convolutional network, which is a forward propagation process. The forward propagation process generates the original private key, and the generated key is used to calculate the total loss $L$ to measure the difference between the currently generated private key and the target one in the transformation domain. In addition, the back propagation algorithm is used to transfer the total loss back to the convolutional layers. The gradient descent method is adopted to implement the back propagation to further update the parameters in each layer to attain better performance, which can be described as:

\begin{equation}
W_{n,i}^j = W_{n,i}^{j{\rm{ - 1}}} - \alpha {\rm{\cdot}}\nabla J(W_n^j)
\end{equation}
In the above equation, $W_{n,i}^j$ represents the values of the parameters ${{w_{n,i}}}$ in the $J$th training round, $\alpha $ represents the learning rate, and $\nabla J(W_n^j)$ represents the gradient of the loss which is passed back to the $n$th convolutional layer in the $j$th training round. The Gradient descent method has the capability to further update the parameters of the network to better learn the mapping relationship. The generator network G and discriminator network D are trained in an alternative manner. When  reaching the setting number of the training epoch or the loss becomes  stabilized, the key is generated as same as the target one in transformation domain. The proposed key generation process is visually shown in Fig.\ref{fig5} and described in Alg. \ref{alg1}.

\renewcommand{\algorithmicrequire}{\textbf{Initialization:}} 
\renewcommand{\algorithmicensure}{\textbf{Output:}} 
\begin{algorithm}[ht]
	\caption{Key Generation Process.}
	\label{alg1}
	\begin{algorithmic}[1]
		\small{
			\Require
			Randomly initialize the parameters $W$ of the DeepKeyGen, ${W_n} = random\left[ {{w_{n,1}},{w_{n,2}},...,{w_{n,i}}} \right]$.
			Define the dimension of $KEY$ as ${\rm{256}} \times {\rm{256}} \times {\rm{3}}$.
			\While {$Epoch{\rm{  <  }}Epoch{_{target}}$}
			\State $x=Convert({IMAGE_{source domain}})$; $y=Convert$
			$({IMAGE_{transformation domain}})$ \verb|//| Convert training images into ${\rm{256}} \times {\rm{256}} \times {\rm{3}}$ matrices.
			\State $KEY=G(x)$ \verb|//| Forward propagation of generator network G. At the last layer of G, the $KEY$ is generated. 
			\State $Result=D(y)$ \verb|//| Forward propagation of discriminator network D. Output the judgment result whether the input is from transformation domain.
			\State $L = L_G + L_D$ \verb|//| Calculate the total loss $L$.
			\State $Backward(L)$ \verb|//| Backward propagation.
			\State $W_{n,i}^j = W_{n,i}^{j{\rm{ - 1}}} - \alpha {\rm{\cdot}}\nabla J(W_n^j)$ \verb|//| Calculate the gradient that pass back to each layer, and then update the network parameters.
			\EndWhile
			\Ensure
			$KEY$ is generated, which is close to the transformation domain.
		}
	\end{algorithmic}
\end{algorithm}

\subsection{Imitation Learning Attack Models}
\label{sec-attack}
In the proposed DeepKeyGen, the structure of the network and the transformation domain are the most important factors in implementing the private key generation. If the structure of the network and the transformation domain are leaked, the attacker may design a similar deep learning network to crack the ciphertext image encrypted by the key generated by the DeepKeyGen. Such an attack is referred to as the ``imitation learning attack'', and the models learned by imitation are known as the ``attack models''. There are three possible scenarios for imitation learning attack, namely: transformation domain leakage, network structure leakage, and transformation domain and network structure leakage (see Sections \ref{subsubsection:Transformation Domain Leakage} to \ref{subsubsection:Transformation Domain and Network Structure Leakage}.

\subsubsection{Transformation Domain Leakage} \label{subsubsection:Transformation Domain Leakage}

This type of leakage assumes that the attacker only knows the transformation domain and adopts the known domain to train the attacking network with different network structures (because the network structure is still in privacy mode) so as to generate attacking private keys to crack ciphertext image. In this scenario, there are two key generator networks with different network structures, key generation network A and key generation network B. The generation network B is regarded as the attack model. Both networks are trained with the same transformation domain. If the private key generated by the network B can be used to decrypt the ciphertext image encrypted with the private key from network A, it means that the attacker cracks the private key through imitation learning.

\subsubsection{Network Structure Leakage}

This attacking scenario assumes that only the network structure of the DeepKeyGen is exposed, and the transformation domain is still in privacy. The attacker adopts different transformation domains to train the network with the same network structure. If the key generated by the attack model can decrypt the ciphertext image, it can be said that the attacker successfully cracks the key.

\subsubsection{Transformation Domain and Network Structure Leakage} \label{subsubsection:Transformation Domain and Network Structure Leakage}

This is the worst scenario, where both the transformation domain and the network structure are leaked. The attacker adopts the same transformation domain and network structure to train the key generation network. If you want to resist this attack, the key generated by the network should be different for each time. It means that the proposed method is asked to perform like a one-time pad method.

\begin{table*}[]
	\caption{The details of generator network G.}
	\label{tab1}
	\begin{center}
		\begin{tabular}{|c|c|c|c|c|c|c|}
			\hline
			\textbf{Convolution Layer}& \textbf{Number}& \textbf{Size}& \textbf{Input Channels}& \textbf{Output Channels}& \textbf{Parameters}& \textbf{Total Parameters}\\
			\hline
			{\color[HTML]{000000} Down Convolution1}                                                        & {\color[HTML]{000000} 1}                                                            & {\color[HTML]{000000} 7*7}                                                        & {\color[HTML]{000000} 3}                                                                    & {\color[HTML]{000000} 16}                                                                    & {\color[HTML]{000000} 2352}                                                             & {\color[HTML]{000000} 2352}                                                                   \\ \hline
			{\color[HTML]{000000} Down Convolution2}                                                        & {\color[HTML]{000000} 1}                                                            & {\color[HTML]{000000} 3*3}                                                        & {\color[HTML]{000000} 16}                                                                   & {\color[HTML]{000000} 32}                                                                    & {\color[HTML]{000000} 4608}                                                             & {\color[HTML]{000000} 6960}                                                                   \\ \hline
			{\color[HTML]{000000} Down Convolution3}                                                        & {\color[HTML]{000000} 1}                                                            & {\color[HTML]{000000} 3*3}                                                        & {\color[HTML]{000000} 32}                                                                   & {\color[HTML]{000000} 64}                                                                    & {\color[HTML]{000000} 18432}                                                            & {\color[HTML]{000000} 25392}                                                                  \\ \hline
			{\color[HTML]{000000} Residual block}                                                           & {\color[HTML]{000000} 6}                                                            & {\color[HTML]{000000} 3*3}                                                        & {\color[HTML]{000000} 64}                                                                   & {\color[HTML]{000000} 64}                                                                    & {\color[HTML]{000000} 221184}                                                           & {\color[HTML]{000000} 246576}                                                                 \\ \hline
			{\color[HTML]{000000} Up Convolution1}                                                          & {\color[HTML]{000000} 1}                                                            & {\color[HTML]{000000} 3*3}                                                        & {\color[HTML]{000000} 64}                                                                   & {\color[HTML]{000000} 32}                                                                    & {\color[HTML]{000000} 18432}                                                            & {\color[HTML]{000000} 265008}                                                                 \\ \hline
			{\color[HTML]{000000} Up Convolution2}                                                          & {\color[HTML]{000000} 1}                                                            & {\color[HTML]{000000} 3*3}                                                        & {\color[HTML]{000000} 32}                                                                   & {\color[HTML]{000000} 16}                                                                    & {\color[HTML]{000000} 4608}                                                             & {\color[HTML]{000000} 269616}                                                                 \\ \hline
			{\color[HTML]{000000} Up Convolution3}                                                          & {\color[HTML]{000000} 1}                                                            & {\color[HTML]{000000} 7*7}                                                        & {\color[HTML]{000000} 16}                                                                   & {\color[HTML]{000000} 3}                                                                     & {\color[HTML]{000000} 2352}                                                             & {\color[HTML]{000000} 198240}                                                                 \\ \hline
		\end{tabular}
	\end{center}
\end{table*}

\begin{table*}[]
	\caption{The details of discriminator network D.}
	\label{tab2}
	\begin{center}
		\begin{tabular}{|c|c|c|c|c|c|c|}
			\hline
			\textbf{Convolution Layer}& \textbf{Number}& \textbf{Size}& \textbf{Input Channels}& \textbf{Output Channels}& \textbf{Parameters}& \textbf{Total Parameters}\\
			\hline
			{\color[HTML]{000000} State1}                                                                   & {\color[HTML]{000000} 1}                                                            & {\color[HTML]{000000} 4*4}                                                        & {\color[HTML]{000000} 3}                                                                    & {\color[HTML]{000000} 16}                                                                    & {\color[HTML]{000000} 768}                                                              & {\color[HTML]{000000} 768}                                                                    \\ \hline
			{\color[HTML]{000000} State2}                                                                   & {\color[HTML]{000000} 1}                                                            & {\color[HTML]{000000} 4*4}                                                        & {\color[HTML]{000000} 16}                                                                   & {\color[HTML]{000000} 32}                                                                    & {\color[HTML]{000000} 8192}                                                             & {\color[HTML]{000000} 8960}                                                                   \\ \hline
			{\color[HTML]{000000} State3}                                                                   & {\color[HTML]{000000} 1}                                                            & {\color[HTML]{000000} 4*4}                                                        & {\color[HTML]{000000} 32}                                                                   & {\color[HTML]{000000} 64}                                                                    & {\color[HTML]{000000} 32768}                                                            & {\color[HTML]{000000} 41728}                                                                  \\ \hline
			{\color[HTML]{000000} State4}                                                                   & {\color[HTML]{000000} 1}                                                            & {\color[HTML]{000000} 4*4}                                                        & {\color[HTML]{000000} 64}                                                                   & {\color[HTML]{000000} 128}                                                                   & {\color[HTML]{000000} 131072}                                                           & {\color[HTML]{000000} 172800}                                                                 \\ \hline
			{\color[HTML]{000000} State5}                                                                   & {\color[HTML]{000000} 1}                                                            & {\color[HTML]{000000} 4*4+1}                                                      & {\color[HTML]{000000} 128}                                                                  & {\color[HTML]{000000} 128}                                                                   & {\color[HTML]{000000} 278528}                                                           & {\color[HTML]{000000} 451328}                                                                 \\ \hline
		\end{tabular}
	\end{center}
\end{table*}

\section{Security Analysis}

In this paper, the bit-wise XOR algorithm is adopted as an encryption and decryption algorithm for evaluating the proposed DeepKeyGen. We will show that the proposed key generation method can achieve high security level in medical images, even with a simple encryption algorithm (e.g., XOR algorithm). In our evaluations, the source domain uses the images from Montgomery County's chest X-ray dataset, and the transformation domain will be the images encrypted using the traditional chaotic system. The chest image from Montgomery County's chest X-ray dataset is adopted as the ``seed'' to generate the private key. There are 138 images in the source domain and the transformation domain. One can observe that the proposed method only requires a small number of images during the training process.

The network structures are shown in Tables \ref{tab1} and \ref{tab2}. During training, the resolution of the input image is ${\rm{256}} \times {\rm{256}}$. All the weight parameters of the network are randomly initialized, and the batch size is set to 1. The Adam optimizer is adopted to optimize the loss function. The initial value of the learning rate is 0.0002. The exponential decay rate of the first-order moment estimation is 0.5, and the second-order moment estimation is 0.999. The training epoch is set to 20000 to achieve a better performance. 

We use three datasets in the evaluation (i.e., Montgomery County's chest X-ray dataset \cite{ds1}, the Ultrasonic Brachial Plexus dataset \cite{ds2}, and the BraTS18 dataset \cite{ds3,ds4}), as they represent three different anatomical parts. The images from these three datasets are also collected from three different medical imaging devices, which are the most commonly used in clinical practice. All experiments run on Nvidia GTX 2080Ti.

\subsection{Key Security Analysis}

\subsubsection{Key Space}
The key space size dictates the resiliency to exhaustive attacks. In the proposed DeepKeyGen, the generated private key is in the form of an image. The dimension of the image is ${\rm{256}} \times {\rm{256}} \times {\rm{3}}$, and the value of each pixel is from 0 to 255. Therefore, the key space of the generated key is ${{\rm{(}}{{\rm{2}}^{\rm{8}}})^{196608}}$. Consequently, this significantly raises the challenge of  attackers to correctly guess the private key and the key space is sufficiently large to resist exhaustive attacks.

\subsubsection{One-Time Pad}
\label{sec-otp}
As the training process of the deep learning network is extremely random, the generated private key will differ for every training and has high randomness. In other words, even if the network structure, the source domain, the transformation domain and all settings remain constant / unchanged, DeepKeyGen will not be able to generate the same private key after being trained at different times. 

\begin{figure}[htbp]
	\centering
	\subfigure{
		\includegraphics[width=0.10\textwidth]{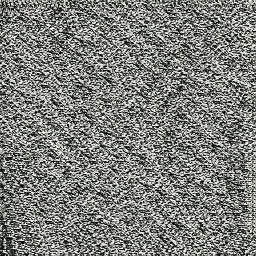} 
	}
	\subfigure{
		\includegraphics[width=0.10\textwidth]{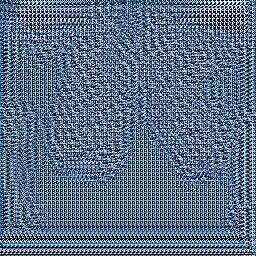} 
	}
	\subfigure{
		\includegraphics[width=0.10\textwidth]{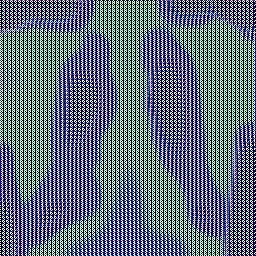} 
	}
	\subfigure{
		\includegraphics[width=0.10\textwidth]{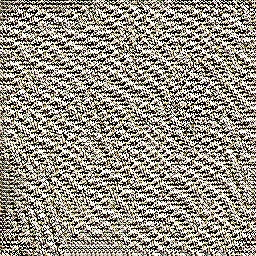} 
	}
	\caption{Visualization of the generated private keys, where the proposed DeepKeyGen is trained four times respectively under the same experimental conditions to generate these four private keys.}
	\label{fig6}
\end{figure}

In our evaluation, the DeepKeyGen is trained four times respectively with the same training settings, and four DeepKeyGens with different weights of network are obtained. Then, the same image (``seed'') is used as the input for these four DeepKeyGens to generate four private keys. As observed from the content, color and contour of the images in Fig.\ref{fig6}, these four private keys differ visually. In other words, the proposed DeepKeyGen can be considered as a type of one-time pad.

\subsubsection{Information Entropy}
Information entropy is used to indicate the degree of uncertainty for a system, and to evaluate the randomness of the private key. The information entropy is defined as:

\begin{equation}
H(m) = \sum\limits_{i = 0}^{{2^n} - 1} {p({m_i})} {\log _2}\frac{1}{{p({m_i})}}
\label{eq0}
\end{equation}
In the above equation, $p({m_i})$ represents the probability of symbol m. The maximum value of entropy is 8 for the grayscale images. Table \ref{tab03} shows the entropy of eight private keys. It can be seen that the entropy of the private key is mostly around 7.98, which indicates that the generated private key has high randomness (i.e., this implies the security of the private keys).

\begin{table}[htbp]
	\small
	\caption{Information entropy of keys.}
	\label{tab03}
	\setlength{\tabcolsep}{0.3mm}{
		\begin{center}			
			\begin{tabular}{|c|c|c|c|c|c|c|c|c|}
				\hline
				\textbf{KEY ID}& \textbf{1}& \textbf{2}& \textbf{3}& \textbf{4}& \textbf{5}& \textbf{6}& \textbf{7}& \textbf{8}\\
				\hline
				{\color[HTML]{000000} Entropy} &
				{\color[HTML]{000000} 7.9807} &
				{\color[HTML]{000000} 7.9801} &
				{\color[HTML]{000000} 7.9808} &
				{\color[HTML]{000000} 7.9806} &
				{\color[HTML]{000000} 7.9790} &
				{\color[HTML]{000000} 7.9782} &
				{\color[HTML]{000000} 7.9791} &
				{\color[HTML]{000000} 7.9803} \\ \hline
			\end{tabular}%
		\end{center}
	}
\end{table}

\subsubsection{Histogram Analysis}

Histogram shows the distribution of the pixel gray values of an image. As the key generated by DeepKeyGen is in the form of the image, the histogram is used to evaluate the distribution of ${p_i}$ for the private key. The horizontal axis of the histogram represents the value of ${p_i}$ and the range is from 0 to 255. The vertical axis of the histogram represents the frequency of each pixel value. The sum of the frequency for all pixel values is equal to one. 

\begin{figure}[htbp]
	\centering
	\subfigure{
		\includegraphics[width=0.10\textwidth]{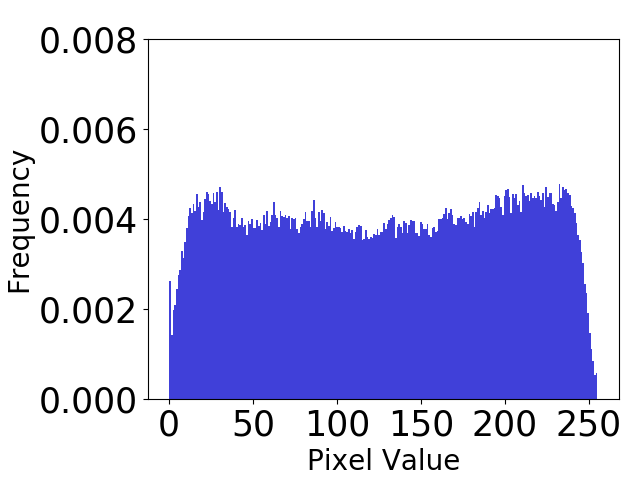}
	}
	\subfigure{
		\includegraphics[width=0.10\textwidth]{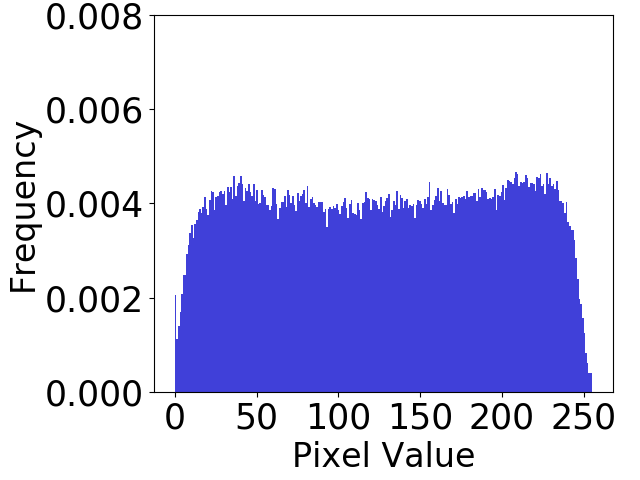}
	}
	\subfigure{
		\includegraphics[width=0.10\textwidth]{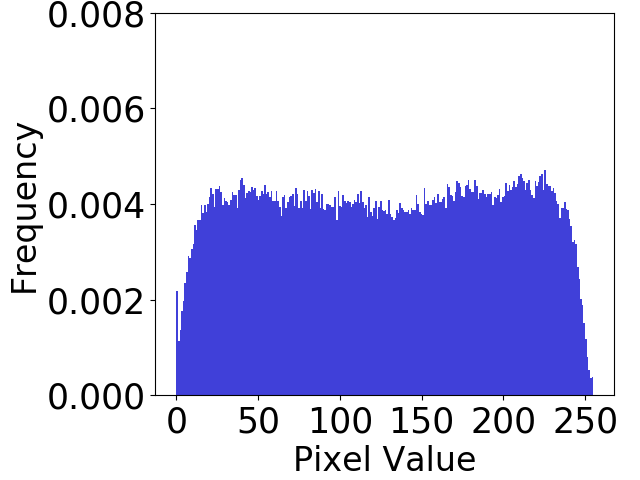}
	}
	\subfigure{
		\includegraphics[width=0.10\textwidth]{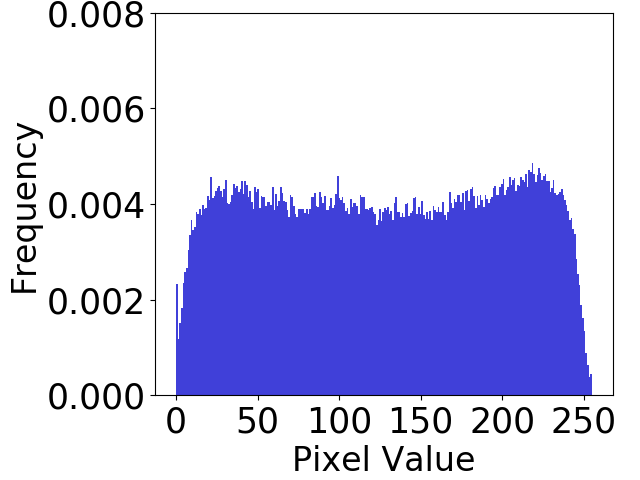}
	}
	\caption{Histograms of the keys generated by DeepKeyGen.}
	\label{fig7}
\end{figure}

The histogram of one key is shown in Fig.\ref{fig7}. It can be seen that the histogram of the private key is relatively uniform. It means that the distribution of the ${p_i}$ is uniform and the frequency of the ${p_i}$ is close. This implies that the generated private key has high randomness and  can resist statistical attacks.

\subsubsection{Sensitivity Analysis}

A stream cipher generator with high security is sensitive to the ``seed''. For DeepKeyGen, the initial image is the ``seed'' adopted to generate the private key. In the evaluation, one-pixel value of the initial image  is randomly changed. Next, the two  images before and after changing one-pixel value are used as inputs  into the DeepKeyGen to generate two  private keys. Then, the differences between two private keys  are calculated using these  two metrics, in order to evaluate the sensitivity to the ``seed''.  

Two metrics, the Number of Pixel Change Rate (NPCR) and the Unified Average Changing Intensity (UACI) are adopted to evaluate the differences between the private keys. NPCR denotes the pixel change rate, which is used to indicate the ratio of different pixel values at the same location of two images. The definition of NPCR is as follows:

\begin{equation}
NPCR = \frac{\sum_{i=0}^W \sum_{j=0}^H D(i, j)}{W \times H} \times 100\%
\label{eq}
\end{equation}	

UACI denotes the intensity of normalized average change, which is used to indicate the average changed density of two images. The definition of UACI is as follows:

\begin{equation}
UACI = \frac{{\sum\limits_{i = 1}^W {\sum\limits_{j = 1}^H {|{T_1}(i,j) - {T_2}(i,j)|} } }}{{255 \times W \times H}} \times 100\%	
\label{eq1}
\end{equation}	
In the above equation, W and H are the width and the height of the image respectively. T(i, j) represents the pixel value in the position (i, j) of the image. If $T_1(i,j) = T_2(i,j)$, then $D(i, j) = 1$. Otherwise (i.e., $T_1(i,j) \neq T_2(i,j)$), $D(i, j) = 0$. 

As shown in Table \ref{tab13}, the results indicate that a small change to the initial image (only change one-pixel value) can result in over 99.5\% differences  between two generated private keys and the average changes in intensity is more than 20\%. This shows that the private key generated by the proposed DeepKeyGen is  sensitive to the ``seed''; thus, the generated private key achieves both pseudo-randomness and uncertainty.

\subsection{Ciphertext Security Analysis}

\subsubsection{Information Entropy Analysis}

If the encrypted image has sufficient randomness, then its information entropy should be very close to 8. Tables \ref{tab3}, \ref{tab4} and \ref{tab5} present the before and after encryption information entropy of eight medical plaintext images for the X-ray dataset, ultrasonic brachial plexus dataset and BraTS18 dataset, respectively. It can be found that the average information entropy of the original chest images, brachial plexus image and the brain image is 7.7360, 7.3878 and 5.8041 respectively. While the average information entropy of the corresponding ciphertext images is respectively 7.9986, 7.9969 and 7.9939, the information entropy of the ciphertext images has significantly improved in comparison to the plaintext images. Furthermore, the information entropy of these encrypted image is very close to the theoretical optimal value of 8. This implies that the ciphertext image has high randomness, where the private key generated using DeepKeyGen is used to encrypt the medical image, and the statistical information of the original image is successfully protected from statistical attacks.

\begin{table}[H]
	\small
	\centering
	\caption{NPCR and UACI calculated by two private keys.}
	\label{tab13}
	\setlength{\tabcolsep}{0.5mm}{
		\begin{tabular}{|c|c|c|c|c|c|c|c|c|}
			\hline
			\textbf{KEYS ID}& \textbf{1}& \textbf{2}& \textbf{3}& \textbf{4}& \textbf{5}& \textbf{6}& \textbf{7}& \textbf{8}\\
			\hline
			{\color[HTML]{000000} NPCR(\%)} &
			{\color[HTML]{000000} 99.56} &
			{\color[HTML]{000000} 99.58} &
			{\color[HTML]{000000} 99.58} &
			{\color[HTML]{000000} 99.56} &
			{\color[HTML]{000000} 99.57} &
			{\color[HTML]{000000} 99.59} &
			{\color[HTML]{000000} 99.63} &
			{\color[HTML]{000000} 99.59} \\ \hline
			{\color[HTML]{000000} UACI(\%)} &
			{\color[HTML]{000000} 20.87} &
			{\color[HTML]{000000} 21.45} &
			{\color[HTML]{000000} 20.31} &
			{\color[HTML]{000000} 23.20} &
			{\color[HTML]{000000} 21.84} &
			{\color[HTML]{000000} 20.36} &
			{\color[HTML]{000000} 20.97} &
			{\color[HTML]{000000} 22.67} \\ \hline
		\end{tabular}%
	}
\end{table}

\begin{table}[htbp]
	\footnotesize
	\centering
	\caption{Information entropy analysis on Montgomery County chest X-ray dataset.}
	\label{tab3}
	\setlength{\tabcolsep}{0.4mm}{
		\begin{tabular}{|c|c|c|c|c|c|c|c|c|}
			\hline
			\textbf{IMAGE ID}& \textbf{1}& \textbf{2}& \textbf{3}& \textbf{4}& \textbf{5}& \textbf{6}& \textbf{7}& \textbf{8}\\
			\hline
			{\color[HTML]{000000} Plain Image} &
			{\color[HTML]{000000} 7.7756} &
			{\color[HTML]{000000} 7.5715} &
			{\color[HTML]{000000} 7.7604} &
			{\color[HTML]{000000} 7.7585} &
			{\color[HTML]{000000} 7.7841} &
			{\color[HTML]{000000} 7.8343} &
			{\color[HTML]{000000} 7.7558} &
			{\color[HTML]{000000} 7.6479} \\ \hline
			{\color[HTML]{000000} Cipher Image} &
			{\color[HTML]{000000} 7.9985} &
			{\color[HTML]{000000} 7.9987} &
			{\color[HTML]{000000} 7.9987} &
			{\color[HTML]{000000} 7.9986} &
			{\color[HTML]{000000} 7.9986} &
			{\color[HTML]{000000} 7.9986} &
			{\color[HTML]{000000} 7.9986} &
			{\color[HTML]{000000} 7.9988} \\ \hline
		\end{tabular}%
	}
\end{table}

\begin{table}[htbp]
	\footnotesize
	\centering
	\caption{Information entropy analysis on Ultrasonic Brachial Plexus dataset.}
	\label{tab4}
	\setlength{\tabcolsep}{0.4mm}{
		\begin{tabular}{|c|c|c|c|c|c|c|c|c|}
			\hline
			\textbf{IMAGE ID}& \textbf{1}& \textbf{2}& \textbf{3}& \textbf{4}& \textbf{5}& \textbf{6}& \textbf{7}& \textbf{8}\\
			\hline
			{\color[HTML]{000000} Plain Image} &
			{\color[HTML]{000000} 7.5874} &
			{\color[HTML]{000000} 7.3432} &
			{\color[HTML]{000000} 7.3946} &
			{\color[HTML]{000000} 7.3469} &
			{\color[HTML]{000000} 7.4049} &
			{\color[HTML]{000000} 7.4207} &
			{\color[HTML]{000000} 7.1624} &
			{\color[HTML]{000000} 7.4423} \\ \hline
			{\color[HTML]{000000} Cipher Image} &
			{\color[HTML]{000000} 7.9974} &
			{\color[HTML]{000000} 7.9973} &
			{\color[HTML]{000000} 7.9970} &
			{\color[HTML]{000000} 7.9969} &
			{\color[HTML]{000000} 7.9965} &
			{\color[HTML]{000000} 7.9970} &
			{\color[HTML]{000000} 7.9963} &
			{\color[HTML]{000000} 7.9970} \\ \hline
		\end{tabular}%
	}
\end{table}

\begin{table}[htbp]
	\footnotesize
	\centering
	\caption{Information entropy analysis on BraTS18 dataset.}
	\label{tab5}
	\setlength{\tabcolsep}{0.4mm}{
		\begin{tabular}{|c|c|c|c|c|c|c|c|c|}
			\hline
			\textbf{IMAGE ID}& \textbf{1}& \textbf{2}& \textbf{3}& \textbf{4}& \textbf{5}& \textbf{6}& \textbf{7}& \textbf{8}\\
			\hline
			{\color[HTML]{000000} Plain Image} &
			{\color[HTML]{000000} 5.9737} &
			{\color[HTML]{000000} 6.0054} &
			{\color[HTML]{000000} 5.8111} &
			{\color[HTML]{000000} 5.2625} &
			{\color[HTML]{000000} 6.3590} &
			{\color[HTML]{000000} 5.8324} &
			{\color[HTML]{000000} 5.9459} &
			{\color[HTML]{000000} 5.2428} \\ \hline
			{\color[HTML]{000000} Cipher Image} &
			{\color[HTML]{000000} 7.9942} &
			{\color[HTML]{000000} 7.9948} &
			{\color[HTML]{000000} 7.9929} &
			{\color[HTML]{000000} 7.9934} &
			{\color[HTML]{000000} 7.9957} &
			{\color[HTML]{000000} 7.9935} &
			{\color[HTML]{000000} 7.9934} &
			{\color[HTML]{000000} 7.9931} \\ \hline
		\end{tabular}%
	}
\end{table}

\subsubsection{Histogram Analysis}

The histogram of the plaintext image is usually relatively concentrated. If these statistics patterns can't be eliminated through the encryption process, the attacker has the ability to crack the original image information by using the statistical attacks.

If the encrypted image is with a uniform histogram distribution, it means that the statistics of the plaintext are eliminated, which is conducive to mitigate the statistical analysis. On the contrary, if the histogram of the encrypted image still keeps patterns to follow, the security of the encrypted image is not enough to effectively resist statistical attacks.

\begin{figure}[htbp]
	\centering
	\subfigure[]{
		\includegraphics[width=0.45\textwidth]{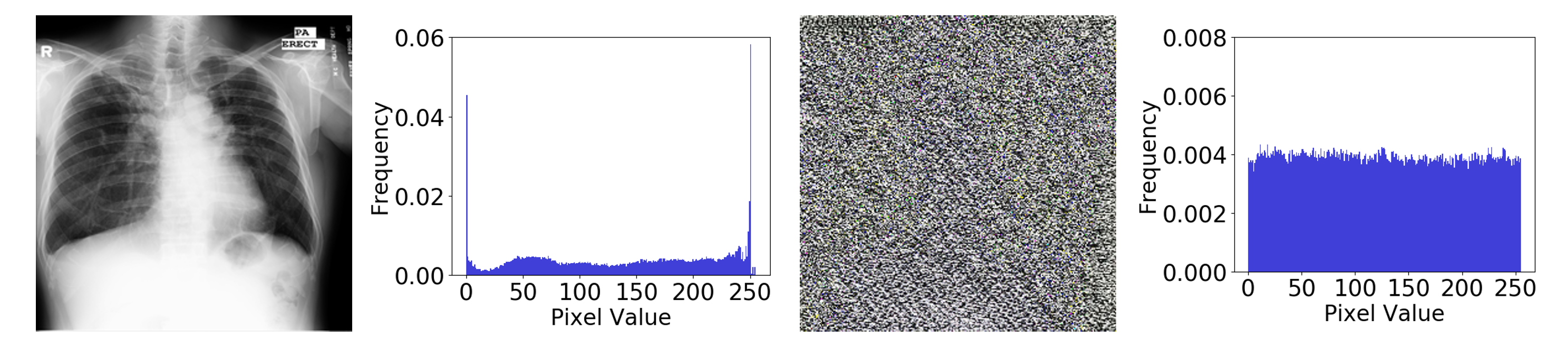} 
	}
	\subfigure[]{
		\includegraphics[width=0.45\textwidth]{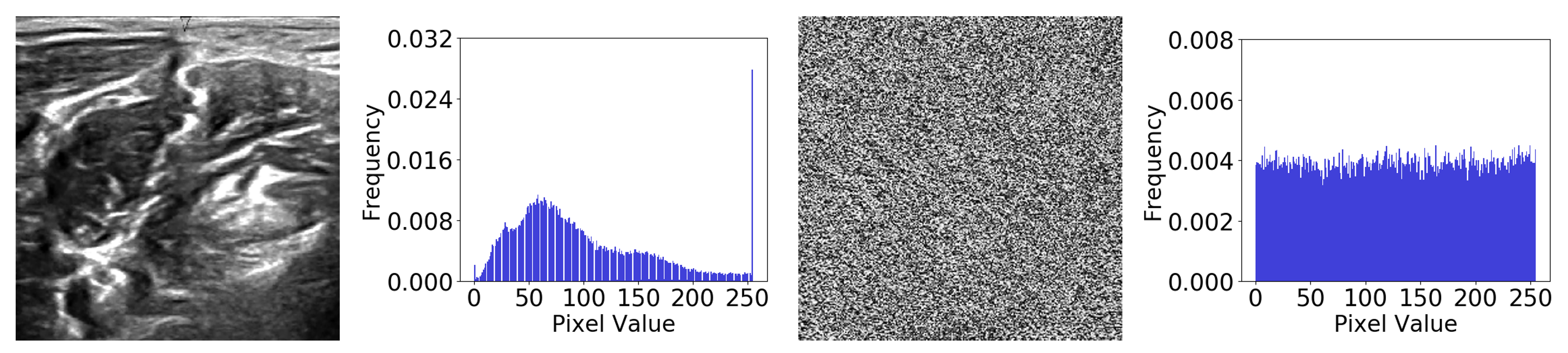} 
	}
	\subfigure[]{
		\includegraphics[width=0.45\textwidth]{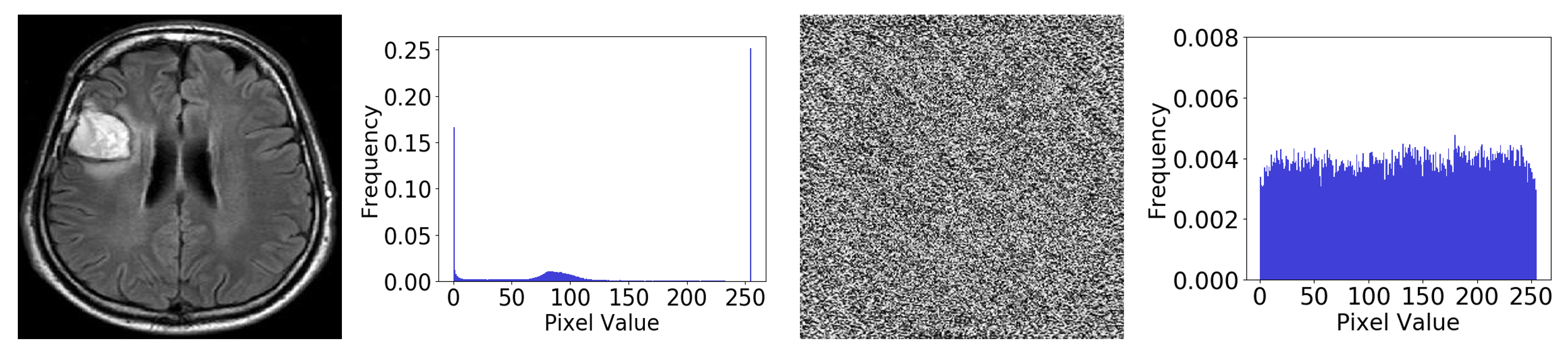} 
	}
	\caption{Histogram analysis of plaintext images and corresponding ciphertext images on (a) Montgomery County chest X-ray dataset, (b) Ultrasonic Brachial Plexus dataset and (c) BraTS18 dataset respectively. On each line, from left to right are the plaintext image, the histogram of plaintext image, the corresponding ciphertext image and the histogram of ciphertext image respectively.}
	\label{fig8}
\end{figure}

In Fig.\ref{fig8} (a), (b) and (c), the original medical images, the histogram of plaintext images, the corresponding ciphertext images and the histograms of ciphertext image are presented from the left to right on Montgomery County chest X-ray dataset, Ultrasonic Brachial Plexus dataset and BraTS18 dataset respectively. It can be seen that the pixel value distributions of the ciphertext images are relatively uniform, which are totally different from the histogram of plaintext images. These distributions are very close to the histogram distribution of white noise, which means the encrypted image can successfully protect the statistical information of the plaintext images. Moreover, it can be proved that the generated private key can facilitate the encryption process for medical image and encrypt the medical image in a random way. Accordingly, it becomes very difficult for the attackers to acquire useful information from these encrypted images by employing the statistical attacks.

\subsubsection{Similarity Analysis}

If the encryption method is with high security, the similarity between the images before and after encryption is quite low. Two metrics mean square error (MSE) and structural similarity (SSIM) are adopted to evaluate the similarity between plaintext and ciphertext images in this experiment. The MSE is defined as:

\begin{equation}
{MSE{\rm{ = }}\frac{{\rm{1}}}{{H \times W \times C}}\sum\limits_{i = 1}^M {\sum\limits_{j = 1}^N {\sum\limits_{k = 1}^C {{{(a(i,j,k) - b(i,j,k))}^2}} } }}
\label{eq0}
\end{equation}
where $H \times W$ is the size of the image, C is the RGB channel of the image, a is the plaintext image and b is the ciphertext image. The larger value of the MSE represents the low similarity between two images. While the SSIM is defined as:

\begin{equation}
SSIM(x,y) = \frac{{(2{\mu _x}{\mu _y} + {c_1})(2{\sigma _{xy}} + {c_2})}}{{(\mu _x^2 + \mu _y^2 + {c_1})(\sigma _x^2 + \sigma _y^2 + {c_2})}}
\label{eq0}
\end{equation}
where x and y are two images, ${{\mu _x}}$ is the mean value of x, ${{\mu _y}}$ is the mean value of y, ${\sigma _x^2}$ is the variance of x, ${\sigma _y^2}$ is the variance of y, ${{\sigma _{xy}}}$ is the covariance of x and y, and ${{c_1}}$ and ${{c_2}}$ are the constant used to maintain stability. The larger value of SSIM indicates a great similarity between two images, where the value range is from 0 to 1. The experimental results are shown in Table \ref{tab6}, \ref{tab7} and \ref{tab8} on three different datasets respectively. The average MSE between the plaintext and ciphertext images on the x-ray dataset, ultrasonic brachial plexus dataset and BraTS18 dataset is 10577.68, 11946.85 and 9741.57 respectively. While the SSIM is 0.0020, 0.0047 and 0.0021 respectively. It can be found that the value of MSE between the plaintext and ciphertext images are very large, while the values of SSIM are close to 0. It indicates that there is a lower similarity between the plaintext and ciphertext images and the proposed method achieves a better encryption performance.

\begin{table}[htbp]
	\scriptsize
	\centering
	\caption{Evaluation of MSE and SSIM on Montgomery County chest X-ray dataset.}
	\label{tab6}
	\setlength{\tabcolsep}{0.1mm}{
		\begin{tabular}{|c|c|c|c|c|c|c|c|c|}
			\hline
			\textbf{IMAGE ID}& \textbf{1}& \textbf{2}& \textbf{3}& \textbf{4}& \textbf{5}& \textbf{6}& \textbf{7}& \textbf{8}
			\\
			\hline
			{\color[HTML]{000000} MSE} &
			{\color[HTML]{000000} 11017.56} &
			{\color[HTML]{000000} 10032.44} &
			{\color[HTML]{000000} 10224.44} &
			{\color[HTML]{000000} 10295.42} &
			{\color[HTML]{000000} 10432.93} &
			{\color[HTML]{000000} 10770.87} &
			{\color[HTML]{000000} 10391.61} &
			{\color[HTML]{000000} 10139.78}
			\\ \hline
			{\color[HTML]{000000} SSIM} &
			{\color[HTML]{000000} 0.0018} &
			{\color[HTML]{000000} 0.0017} &
			{\color[HTML]{000000} 0.0024} &
			{\color[HTML]{000000} 0.0019} &
			{\color[HTML]{000000} 0.0031} &	
			{\color[HTML]{000000} 0.0023} &
			{\color[HTML]{000000} 0.0012} &
			{\color[HTML]{000000} 0.0019}  
			\\ 
			\hline
		\end{tabular}%
	}
\end{table}

\begin{table}[htbp]
	\scriptsize
	\centering
	\caption{Evaluation of MSE and SSIM on Ultrasonic Brachial Plexus dataset.}
	\label{tab7}
	\setlength{\tabcolsep}{0.1mm}{
		\begin{tabular}{|c|c|c|c|c|c|c|c|c|}
			\hline
			\textbf{IMAGE ID}& \textbf{1}& \textbf{2}& \textbf{3}& \textbf{4}& \textbf{5}& \textbf{6}& \textbf{7}& \textbf{8}
			\\
			\hline
			{\color[HTML]{000000} MSE} &
			{\color[HTML]{000000} 11720.98} &
			{\color[HTML]{000000} 10744.93} &
			{\color[HTML]{000000} 11968.21} &
			{\color[HTML]{000000} 11933.00} &
			{\color[HTML]{000000} 12628.54} &
			{\color[HTML]{000000} 11491.53} &
			{\color[HTML]{000000} 11135.18} &
			{\color[HTML]{000000} 13072.60} 
			\\ 
			\hline
			{\color[HTML]{000000} SSIM} &
			{\color[HTML]{000000} 0.0061} &
			{\color[HTML]{000000} 0.0072} &
			{\color[HTML]{000000} 0.0040} &
			{\color[HTML]{000000} 0.0056} &
			{\color[HTML]{000000} 0.0047} &
			{\color[HTML]{000000} 0.0049} &
			{\color[HTML]{000000} 0.0026} &
			{\color[HTML]{000000} 0.0049} 
			\\ 
			\hline
		\end{tabular}%
	}
\end{table}

\begin{table}[htbp]
	\scriptsize
	\centering
	\caption{Evaluation of MSE and SSIM on BraTS18 dataset.}
	\label{tab8}
	\setlength{\tabcolsep}{0.4mm}{
		\begin{tabular}{|c|c|c|c|c|c|c|c|c|}
			\hline
			\textbf{IMAGE ID}& \textbf{1}& \textbf{2}& \textbf{3}& \textbf{4}& \textbf{5}& \textbf{6}& \textbf{7}& \textbf{8}
			\\
			\hline
			{\color[HTML]{000000} MSE} &
			{\color[HTML]{000000} 10176.00} &
			{\color[HTML]{000000} 10694.46} &
			{\color[HTML]{000000} 10300.70} &
			{\color[HTML]{000000} 9455.74} &
			{\color[HTML]{000000} 10715.22} &
			{\color[HTML]{000000} 9287.06} &
			{\color[HTML]{000000} 9052.20} &
			{\color[HTML]{000000} 8701.21} 
			\\ 
			\hline
			{\color[HTML]{000000} SSIM} &
			{\color[HTML]{000000} 0.0017} &
			{\color[HTML]{000000} 0.0027} &
			{\color[HTML]{000000} 0.0024} &
			{\color[HTML]{000000} 0.0004} &
			{\color[HTML]{000000} 0.0005} &
			{\color[HTML]{000000} 0.0019} &
			{\color[HTML]{000000} 0.0022} &
			{\color[HTML]{000000} 0.0014} 
			\\ 
			\hline
		\end{tabular}%
	}
\end{table}

\subsubsection{Correlation Analysis}

In general, there is a strong correlation between adjacent pixels in a plaintext image, so that the pixel values of the image hold certain regular patterns in the horizontal, vertical and diagonal directions. If the attacker observes these patterns and makes use of the correlation to attack, the security of the image becomes vulnerable. It's necessary to reduce the correlation between adjacent pixels in the ciphertext image as much as possible to protect the original medical image. Furthermore, in order to prove the effectiveness of the proposed key generation method, the correlation between adjacent pixels in the ciphertext image should be kept in a low level. 

Assuming that ${x_i}$ and ${y_i}$ are the grayscale values of two adjacent pixels, the correlation value between the two adjacent pixels is:

\begin{equation}
{r_{xy}} = \frac{{{\mathop{\rm cov}} (x,y)}}{{\sqrt {D(x)} \sqrt {D(y)} }}
\label{eq0}
\end{equation}
where
\begin{equation}
{\mathop{\rm cov}} (x,y) = \frac{1}{N}\sum\limits_{i = 1}^N {(({x_i} - E(x))({y_i} - E(y)))} 
\label{eq0}
\end{equation}
\begin{equation}
E(x) = \frac{1}{N}\sum\limits_{i = 1}^N {{x_i}} ,D(x) = \frac{1}{N}\sum\limits_{i = 1}^N {({x_i}}  - E(x){)^2}
\label{eq0}
\end{equation}

For calculating the correlation of the plaintext images and ciphertext images respectively, 256 adjacent pixel points in the horizontal, vertical and diagonal directions of the image are chosen. As shown in Table \ref{tab9}, \ref{tab10} and \ref{tab11}, the symbol $*$ in the table indicates the correlation in the encrypted images. It can be concluded that, for the plaintext images, there is a certain correlation between adjacent pixels. For the images encrypted by the proposed private key, the correlation values between adjacent pixels are almost to zero. It means that the image encrypted by the proposed DeepKeyGen greatly decreases the correlation of pixels in the adjacent pixels of the same position in the corresponding plaintext image. It can effectively resist the attacks by making use of strong correlation. Accordingly, the private key generated by the DeepKeyGen can be regarded as an effective stream cipher to encrypt the medical images.

\begin{table}[htbp]
	\centering
	\caption{Correlation of plaintext images and corresponding ciphertext images respectively in horizontal, vertical and diagonal directions on Montgomery County chest X-ray dataset.}
	\label{tab9}
	\setlength{\tabcolsep}{2mm}{
		\begin{tabular}{|c|c|c|c|c|}
			\hline
			\textbf{IMAGE ID}& \textbf{1}& \textbf{2}& \textbf{3}& \textbf{4}\\
			\hline
			{\color[HTML]{000000} Horizontal} & {\color[HTML]{000000} 0.9636}  & {\color[HTML]{000000} 0.9054} & {\color[HTML]{000000} 0.9421} & {\color[HTML]{000000} 0.9093} \\ \hline
			{\color[HTML]{000000} Horizontal*} &
			{\color[HTML]{000000} 0.0383} &
			{\color[HTML]{000000} 0.0280} &
			{\color[HTML]{000000} 0.0511} &
			{\color[HTML]{000000} 0.0395} \\ \hline
			{\color[HTML]{000000} Vertical}   & {\color[HTML]{000000} 0.7077}  & {\color[HTML]{000000} 0.9927} & {\color[HTML]{000000} 0.7917} & {\color[HTML]{000000} 0.8387} \\ \hline
			{\color[HTML]{000000} Vertical*}  & {\color[HTML]{000000} 0.2259}  & {\color[HTML]{000000} 0.1344} & {\color[HTML]{000000} 0.1785} & {\color[HTML]{000000} 0.1637} \\ \hline
			{\color[HTML]{000000} Diagonal}   & {\color[HTML]{000000} 0.8477}  & {\color[HTML]{000000} 0.8105} & {\color[HTML]{000000} 0.6266} & {\color[HTML]{000000} 0.8407} \\ \hline
			{\color[HTML]{000000} Diagonal*}  & {\color[HTML]{000000} 0.1158} & {\color[HTML]{000000} 0.0380} & {\color[HTML]{000000} 0.0453} & {\color[HTML]{000000} 0.1242} \\ \hline
		\end{tabular}%
	}
\end{table}

\begin{table}[htbp]
	\centering
	\caption{Correlation of plaintext images and corresponding ciphertext images respectively in horizontal, vertical and diagonal directions on Ultrasonic Brachial Plexus dataset.}
	\label{tab10}
	\setlength{\tabcolsep}{2mm}{
		\begin{tabular}{|c|c|c|c|c|}
			\hline
			\textbf{IMAGE ID}& \textbf{1}& \textbf{2}& \textbf{3}& \textbf{4}\\
			\hline
			{\color[HTML]{000000} Horizontal}  & {\color[HTML]{000000} 0.6641}  & {\color[HTML]{000000} 0.8376}  & {\color[HTML]{000000} 0.3959}  & {\color[HTML]{000000} 0.2172}  \\ \hline
			{\color[HTML]{000000} Horizontal*} & {\color[HTML]{000000} 0.0627} & {\color[HTML]{000000} 0.0040}  & {\color[HTML]{000000} 0.0974} & {\color[HTML]{000000} 0.0396} \\ \hline
			{\color[HTML]{000000} Vertical}    & {\color[HTML]{000000} 0.8257}  & {\color[HTML]{000000} 0.7377}  & {\color[HTML]{000000} 0.7285}  & {\color[HTML]{000000} 0.6056}  \\ \hline
			{\color[HTML]{000000} Vertical*}   & {\color[HTML]{000000} 0.2557}  & {\color[HTML]{000000} 0.2938}  & {\color[HTML]{000000} 0.2666}  & {\color[HTML]{000000} 0.2612}  \\ \hline
			{\color[HTML]{000000} Diagonal}    & {\color[HTML]{000000} 0.3933}  & {\color[HTML]{000000} 0.3503}  & {\color[HTML]{000000} 0.6661}  & {\color[HTML]{000000} 0.5839}  \\ \hline
			{\color[HTML]{000000} Diagonal*}   & {\color[HTML]{000000} 0.0203}  & {\color[HTML]{000000} 0.1816} & {\color[HTML]{000000} 0.0399} & {\color[HTML]{000000} 0.0005} \\ \hline
		\end{tabular}%
	}
\end{table}

\begin{table}[htbp]
	\centering
	\caption{Correlation of plaintext images and corresponding ciphertext images respectively in horizontal, vertical and diagonal directions on BraTS18 dataset.}
	\label{tab11}
	\setlength{\tabcolsep}{2mm}{
		\begin{tabular}{|c|c|c|c|c|}
			\hline
			\textbf{IMAGE ID}& \textbf{1}& \textbf{2}& \textbf{3}& \textbf{4}\\
			\hline
			{\color[HTML]{000000} Horizontal}  & {\color[HTML]{000000} 0.8877} & {\color[HTML]{000000} 0.7038} & {\color[HTML]{000000} 0.9387} & {\color[HTML]{000000} 0.8970} \\ \hline
			{\color[HTML]{000000} Horizontal*} & {\color[HTML]{000000} 0.0357} & {\color[HTML]{000000} 0.0724} & {\color[HTML]{000000} 0.1600} & {\color[HTML]{000000} 0.0600}  \\ \hline
			{\color[HTML]{000000} Vertical}    & {\color[HTML]{000000} 0.8456} & {\color[HTML]{000000} 0.4053} & {\color[HTML]{000000} 0.2657} & {\color[HTML]{000000} 0.5725} \\ \hline
			{\color[HTML]{000000} Vertical*}   & {\color[HTML]{000000} 0.2319} & {\color[HTML]{000000} 0.2626} & {\color[HTML]{000000} 0.1966} & {\color[HTML]{000000} 0.2510} \\ \hline
			{\color[HTML]{000000} Diagonal}    & {\color[HTML]{000000} 0.8185} & {\color[HTML]{000000} 0.7037} & {\color[HTML]{000000} 0.7756} & {\color[HTML]{000000} 0.8063} \\ \hline
			{\color[HTML]{000000} Diagonal*}   & {\color[HTML]{000000} 0.1059} & {\color[HTML]{000000} 0.1120} & {\color[HTML]{000000} 0.1130} & {\color[HTML]{000000} 0.0716} \\ \hline
		\end{tabular}%
	}
\end{table}

\subsection{Traditional Attack Models}
\subsubsection{Ciphertext Only Attack}
In such attacks, the attacker only knows the ciphertext image. However, the key space of the generated private key is ${{\rm{(}}{{\rm{2}}^{\rm{8}}})^{196608}}$. In other words, it is challenging for the attacker to correctly guess the generated key and decrypt the image. In addition, the proposed key generation method achieves high randomness and has a complex generation process. Therefore, it is challenging to correctly guess the private key through ciphertext only attacks.

\subsubsection{Known-plaintext Attack}
In a known-plaintext attack, the attacker can obtain a part of the plaintext and the corresponding ciphertext (e.g., intercepting the first part of the information, and obtain the encryption method to facilitate the process of cracking the rest of the corresponding ciphertext). As shown in Tables \ref{tab6} to \ref{tab8}, the similarity between the ciphertext and the plaintext is low across these three datasets. This implies that the attacker is not able to infer the complete plaintext due to the very low smilarity because the plaintext and the corresponding ciphertext. In addition, the information entropy of the ciphertext is relatively high, as shown in Tables \ref{tab3} to \ref{tab5}, which also greatly increases the difficulty of the attack. Moreover, Tables \ref{tab9} to \ref{tab11} show that the correlation in the ciphertext is very low. Consequently, the attacker is not able to infer the complete plaintext based on a contextual analysis of the ciphertext. In other words, DeepKeyGen resists known-plaintext attacks.

\subsubsection{Chosen-plaintext Attack / Differential Attack} \label{subsubsection:Chosen-plaintext Attack / Differential Attack}
In a chosen-plaintext attack, the attacker can observe the change of the ciphertext image by making small changes on the plaintext image, and to indicate the patterns between both plaintext and ciphertext images. Moreover, the differential attack is regarded as a kind of chosen plaintext attack. Recall that DeepKeyGen is designed to generate secure private keys, and the existing XOR algorithm is employed to encrypt/decrypt medical image(s) with the generated private key. Therefore, the chosen plaintext attack is implemented on the private key instead of the plaintext image. To be more specific, our evaluation uses only one-pixel value change on one seed image, and two different private keys are generated using DeepKeyGen. Then, based on one plaintext image, two ciphertext images can be obtained by adopting the XOR algorithm with these two private keys. We use both NPCR and UACI metrics to evaluate the performance. 

In the evaluation, a total of eight plaintext images (eight cases) are used. As shown in Table \ref{tab-diff}, the average NPCR value between two ciphertext images is 99.59\% while the average UACI value is 23.19\%. This implies that even if only one value is changed, there is a major change for the final ciphertext image. Consequently, it would be challenging for an attacker to find the pattern(s) between both plaintext and ciphertext images. In other words, DeepKeyGen resists chosen-plaintext and differential attacks.

\begin{table}[H]
	\small
	\centering
	\caption{NPCR and UACI values for two ciphertext images.}
	\label{tab-diff}
	\setlength{\tabcolsep}{0.5mm}{
		\begin{tabular}{|c|c|c|c|c|c|c|c|c|}
			\hline
			\textbf{IMAGE ID}& \textbf{1}& \textbf{2}& \textbf{3}& \textbf{4}& \textbf{5}& \textbf{6}& \textbf{7}& \textbf{8}\\
			\hline
			{\color[HTML]{000000} NPCR(\%)} &
			{\color[HTML]{000000} 99.63} &
			{\color[HTML]{000000} 99.61} &
			{\color[HTML]{000000} 99.59} &
			{\color[HTML]{000000} 99.60} &
			{\color[HTML]{000000} 99.59} &
			{\color[HTML]{000000} 99.56} &
			{\color[HTML]{000000} 99.58} &
			{\color[HTML]{000000} 99.61} \\ \hline
			{\color[HTML]{000000} UACI(\%)} &
			{\color[HTML]{000000} 23.01} &
			{\color[HTML]{000000} 24.37} &
			{\color[HTML]{000000} 22.90} &
			{\color[HTML]{000000} 22.51} &
			{\color[HTML]{000000} 23.24} &
			{\color[HTML]{000000} 23.12} &
			{\color[HTML]{000000} 23.46} &
			{\color[HTML]{000000} 22.97} \\ \hline
		\end{tabular}%
	}
\end{table}

\subsubsection{Chosen-ciphertext Attack}
In a chosen-ciphertext attack, the attacker can access the decryption device to construct the plaintext corresponding to any ciphertext. Since we focus only on the key generation rather than the encryption/decryption algorithms, here we will show how the private key resist chosen-ciphertext attacks. We assume that the attacker knows the ciphertext image, ``seed'', and the XOR decryption algorithm, and the attacker seeks to construct the same plaintext image. Recall that DeepKeyGen uses two ``seeds'' that are only one-pixel value different to generate two separate private keys. Then, based on the XOR decryption algorithm, two plaintext images can be obtained using these two private keys to decrypt one ciphertext image. Furthermore, the NPCR and UACI metrics are adopted to evaluate the difference(s) between two plaintext images. Similar to Section \ref{subsubsection:Chosen-plaintext Attack / Differential Attack}, a total of eight plaintext images (eight cases) are used in the evaluation. As shown in Table \ref{tab-diff-b}, the average NPCR value between two plaintext images is 99.59\% while the average UACI value is 23.66\%. This implies that the proposed method resists chosen-ciphertext attacks.

\begin{table}[H]
	\small
	\centering
	\caption{NPCR and UACI values for two plaintext images.}
	\label{tab-diff-b}
	\setlength{\tabcolsep}{0.5mm}{
		\begin{tabular}{|c|c|c|c|c|c|c|c|c|}
			\hline
			\textbf{IMAGE ID}& \textbf{1}& \textbf{2}& \textbf{3}& \textbf{4}& \textbf{5}& \textbf{6}& \textbf{7}& \textbf{8}\\
			\hline
			{\color[HTML]{000000} NPCR(\%)} &
			{\color[HTML]{000000} 99.71} &
			{\color[HTML]{000000} 99.57} &
			{\color[HTML]{000000} 99.49} &
			{\color[HTML]{000000} 99.55} &
			{\color[HTML]{000000} 99.64} &
			{\color[HTML]{000000} 99.50} &
			{\color[HTML]{000000} 99.68} &
			{\color[HTML]{000000} 99.61} \\ \hline
			{\color[HTML]{000000} UACI(\%)} &
			{\color[HTML]{000000} 23.90} &
			{\color[HTML]{000000} 24.12} &
			{\color[HTML]{000000} 23.57} &
			{\color[HTML]{000000} 22.95} &
			{\color[HTML]{000000} 24.24} &
			{\color[HTML]{000000} 22.60} &
			{\color[HTML]{000000} 24.07} &
			{\color[HTML]{000000} 23.84} \\ \hline
		\end{tabular}%
	}
\end{table}

\subsection{Security Against Imitation Learning Attackers}
We will evaluate whether the attacker can use imitation learning attacks (see Section~\ref{sec-attack}) to generate an appropriate key that can decrypt the target ciphertext image.

\subsubsection{Transformation Domain Leakage}

For the transformation domain leakage, four different network structures, say network A, network B, network C and network D, are designed as the attacking model. The network structure of these four networks is shown in Table \ref{tab12}. With the exception of the network structure, the transformation domain and all other experimental settings are kept the same to train DeepKeyGen. 

\begin{table}[H]
	\centering
	\caption{Structure of four networks, i.e., networks A to D.}
	\label{tab12}
	\setlength{\tabcolsep}{2mm}{
		\begin{tabular}{|c|c|c|c|c|}
			\hline
			\textbf{Convolution Layer}& \textbf{Net.A}& \textbf{Net.B}& \textbf{Net.C}& \textbf{Net.D}\\
			\hline
			{\color[HTML]{000000} Down Convolution1} & {\color[HTML]{000000} 1} & {\color[HTML]{000000} 1} & {\color[HTML]{000000} 1} & {\color[HTML]{000000} 1}  \\ \hline
			{\color[HTML]{000000} Down Convolution2} & {\color[HTML]{000000} 1} & {\color[HTML]{000000} 1} & {\color[HTML]{000000} 1} & {\color[HTML]{000000} 1}  \\ \hline
			{\color[HTML]{000000} Down Convolution3} & {\color[HTML]{000000} 1} & {\color[HTML]{000000} 1} & {\color[HTML]{000000} 1} & {\color[HTML]{000000} 1}  \\ \hline
			{\color[HTML]{000000} Residual block}    & {\color[HTML]{000000} 3} & {\color[HTML]{000000} 6} & {\color[HTML]{000000} 9} & {\color[HTML]{000000} 12} \\ \hline
			{\color[HTML]{000000} Up Convolution1}   & {\color[HTML]{000000} 1} & {\color[HTML]{000000} 1} & {\color[HTML]{000000} 1} & {\color[HTML]{000000} 1}  \\ \hline
			{\color[HTML]{000000} Up Convolution2}   & {\color[HTML]{000000} 1} & {\color[HTML]{000000} 1} & {\color[HTML]{000000} 1} & {\color[HTML]{000000} 1}  \\ \hline
			{\color[HTML]{000000} Up Convolution3}   & {\color[HTML]{000000} 1} & {\color[HTML]{000000} 1} & {\color[HTML]{000000} 1} & {\color[HTML]{000000} 1}  \\ \hline
		\end{tabular}%
	}
\end{table}

\begin{figure}[H]
	\centering
	\subfigure[]{
		\includegraphics[width=0.06\textwidth]{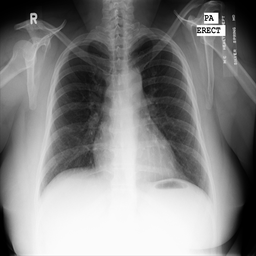} 
	}
	\subfigure[]{
		\includegraphics[width=0.06\textwidth]{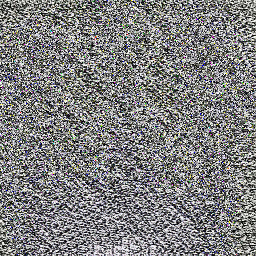} 
	}
	\subfigure[]{
		\includegraphics[width=0.06\textwidth]{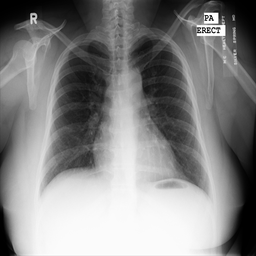} 
	}
	\subfigure[]{
		\includegraphics[width=0.06\textwidth]{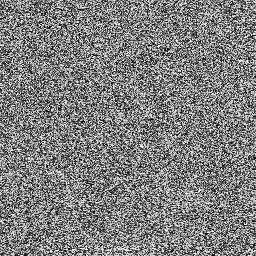} 
	}
	\subfigure[]{
		\includegraphics[width=0.06\textwidth]{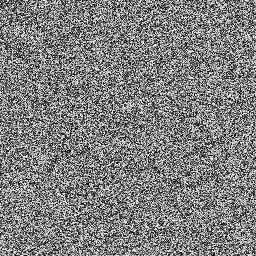} 
	}
	\subfigure[]{
		\includegraphics[width=0.06\textwidth]{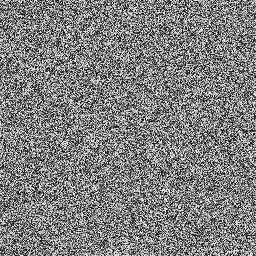} 
	}
	\caption{Attack model analysis of transformation domain leakage. (a) Plaintext image. (b) Ciphertext image encrypted using key A. (c) Plaintext image decrypted using key A. (d) Plaintext image decrypted using key B. (e) Plaintext image decrypted using key C. (f) Plaintext image decrypted using key D.}
	\label{fig11}
\end{figure}

The original plaintext image is encrypted by using a key generated by the trained network A. The ciphertext image is then decrypted by using the keys generated by network A, network B, network C and network D (key A, key B, key C and key D) respectively to obtain the restored original images. As shown in Fig.\ref{fig11}, it can be found that the original image (Fig.\ref{fig11} (a)) encrypted by key A (Fig.\ref{fig11} (b)) can only be correctly decrypted by key A. The images decrypted by key B, key C and key D cannot be visually recognized, and the results are shown in Fig.\ref{fig11} (d), Fig.\ref{fig11} (e) and  Fig.\ref{fig11} (f) respectively. The experiment shows that when the attacker only knows the transformation domain, it is impossible to generate the same private key and decrypt the ciphertext image by constructing an attack model.

\subsubsection{Network Structure Leakage}

\begin{figure}[H]
	\centering
	\subfigure[]{
		\includegraphics[width=0.06\textwidth]{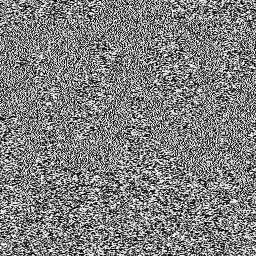} 
	}
	\subfigure[]{
		\includegraphics[width=0.06\textwidth]{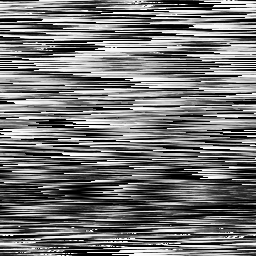} 
	}
	\subfigure[]{
		\includegraphics[width=0.06\textwidth]{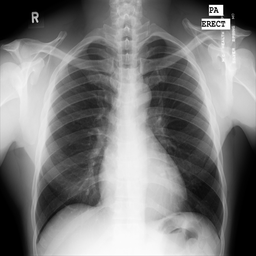} 
	}
	\subfigure[]{
		\includegraphics[width=0.06\textwidth]{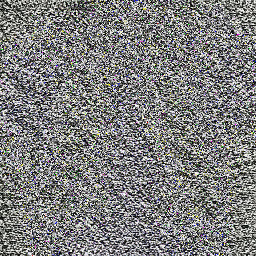} 
	}
	\subfigure[]{
		\includegraphics[width=0.06\textwidth]{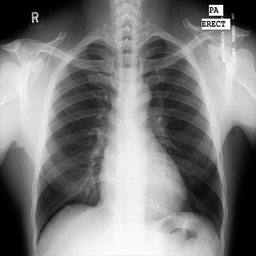} 
	}
	\subfigure[]{
		\includegraphics[width=0.06\textwidth]{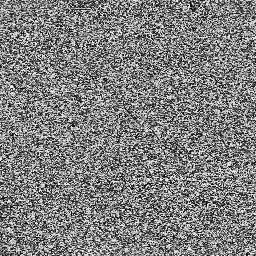} 
	}
	\caption{Attack model analysis of network structure leakage. (a) Transformation domain A. (b) Transformation domain B. c plaintext image. (d) Ciphertext image encrypted using key A. (e) Plaintext image decrypted using key A. (f) Plaintext image decrypted using key B.}
	\label{fig12}
\end{figure}

As shown in Fig.\ref{fig12} (a) and Fig.\ref{fig12} (b), two different transformation domains, transformation domain A and transformation domain B, are adopted to train the DeepKeyGen with the same network structure, while other experimental settings are kept the same. And the private key A and private key B are obtained respectively. 

In Fig.\ref{fig12}, the Fig.\ref{fig12} (c) is the original image and Fig.\ref{fig12} (d) represents the image encrypted by key A. The Fig.\ref{fig12} (e) and Fig.\ref{fig12} (f) are the decrypted images by using the key A and key B respectively. Seen from the experimental results, the key B cannot decrypt the image encrypted by the key A. It can be proved that the generated private keys trained with different transformation domains cannot mutually decrypt the corresponding ciphertext images. It means even if the attacker knows the network structure, it cannot obtain the private key and decrypt the corresponding ciphertext image without knowing the transformation domain.

\subsubsection{Transformation Domain and Network Structure Leakage}

In this scenario, four DeepKeyGens with the same network structure and the same transformation domain are trained to obtain network A, network B, network C and network D respectively. And they also generate the private key A, private key B, private key C and private key D respectively. The experiment evaluates the encryption performance and mutual decryption performance by adopting these four keys on the one same plaintext medical image.

\begin{figure}[H]
	\centering
	\subfigure{
		\includegraphics[width=0.06\textwidth]{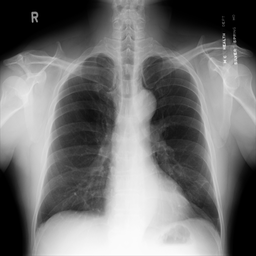} 
	}
	\subfigure{
		\includegraphics[width=0.06\textwidth]{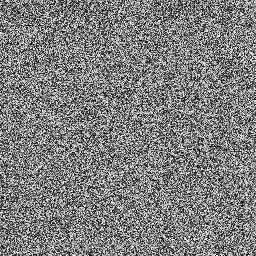} 
	}
	\subfigure{
		\includegraphics[width=0.06\textwidth]{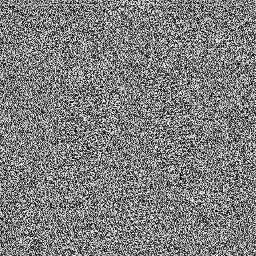} 
	}
	\subfigure{
		\includegraphics[width=0.06\textwidth]{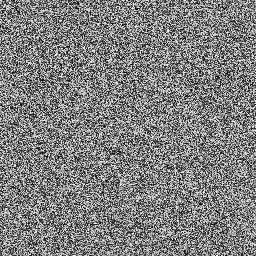} 
	}
	
	\subfigure{
		\includegraphics[width=0.06\textwidth]{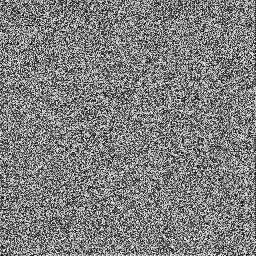} 
	}
	\subfigure{
		\includegraphics[width=0.06\textwidth]{figs/attack/13.png} 
	}
	\subfigure{
		\includegraphics[width=0.06\textwidth]{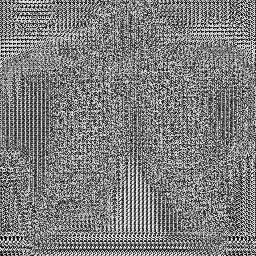} 
	}
	\subfigure{
		\includegraphics[width=0.06\textwidth]{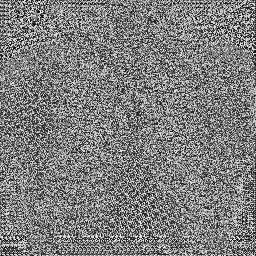} 
	}
	
	\subfigure{
		\includegraphics[width=0.06\textwidth]{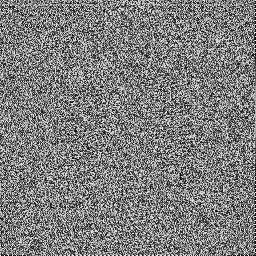} 
	}
	\subfigure{
		\includegraphics[width=0.06\textwidth]{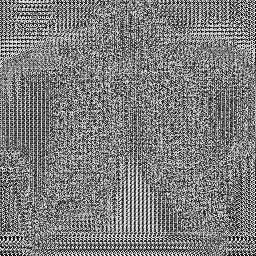} 
	}
	\subfigure{
		\includegraphics[width=0.06\textwidth]{figs/attack/13.png} 
	}
	\subfigure{
		\includegraphics[width=0.06\textwidth]{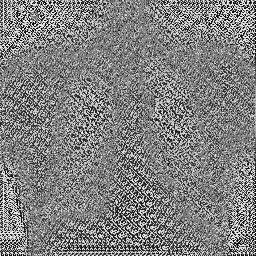} 
	}
	
	\subfigure{
		\includegraphics[width=0.06\textwidth]{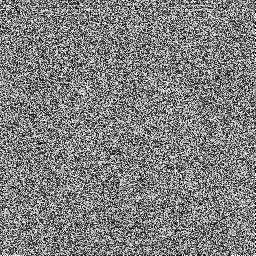} 
	}
	\subfigure{
		\includegraphics[width=0.06\textwidth]{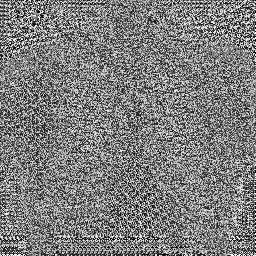} 
	}
	\subfigure{
		\includegraphics[width=0.06\textwidth]{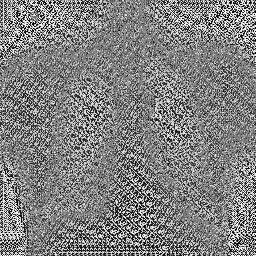} 
	}
	\subfigure{
		\includegraphics[width=0.06\textwidth]{figs/attack/13.png} 
	}
	\caption{The mutual decryption performance of four keys generated under the same experimental conditions.}
	\label{fig13}
\end{figure}

As shown in Fig.\ref{fig13}, the first line is four decrypted images which use four keys to decrypt the images encrypted by private key A, and the rest lines can be deduced by analogy. It can be clearly seen that one key cannot be used to decrypt the images encrypted with other keys. The results of the experiment demonstrate that even if the attacker knows the network structure and transformation domain, it's impossible to obtain the same key generation network, and is unable to generate the same private key to crack the ciphertext image. This experiment can be regarded as a supplementary experiment of one-time pad analysis in Section~\ref{sec-otp}. Furthermore, it can be proven that the proposed key generation network is one type of one-time pad method and can resist the attack model even with the leakage of both the transformation domain and network architecture.

Overall, even if one or both of the transformation domain and network structure of the DeepKeyGen are leaked, the attacker cannot train the same key generation network to generate the same private key. The DeepKeyGen can be regarded as one type of one-time pad method, which guarantees the security of the ciphertext image.

\subsection{Comparison with Existing Key Generation Algorithms}
In this section, some existing key generation algorithms are adopted to compare with the proposed DeepKeyGen algorithm. These algorithms include chaotic system, linear congruential generator (LCG), mersenne twister, rivest cipher 4 (RC4) and RSA algorithm. In order to evaluate the security of generated private key, the experiments are evaluated from the key space, information entropy and the randomness, respectively.

For the key space, we mainly compare the two widely used encryption algorithms, RC4 and RSA are adopted as the comparison. The key space of RC4 and RSA both are ${2^{2048}}$, which is much smaller than the proposed method ${{\rm{(}}{{\rm{2}}^{\rm{8}}})^{196608}}$. It indicates that the proposed DeepKeyGen can better resist the brute force attack.

For the information entropy of private key, the experimental results are shown in Table \ref{tab-com-key-entro}. It can be seen that the values of information entropy for private keys generated by these algorithms are very close, which all are around 7.9900. It can be proven that the private key generated by DeepKeyGen achieves a good performance in term of randomness.

Moreover, the evaluation metric non-overlapping template matching, binary matrix rank, maurer's ``universal statistical'' and random excursions variant are also employed to evaluate the randomness of the generated private key. The non-overlapping template matching is used to detect whether generators produce too many occurrences of a given non-periodic pattern. Binary matrix rank is adopted to measure the linear dependence among fixed length substrings of the original sequence. Maurer's ``universal statistical'' is employed to detect whether the generated sequence can be significantly compressed without information loss. The random excursions variant is used to detect the deviations from the expected number of visits to various states in the random walk. These four metrics are indicated by the P-value. If the P-value is $\geq 0.01$, it represents that the private key is with high randomness. In order to keep the fairness of the experiment, the same chest image from Montgomery County chest X-ray dataset is employed as the ``seed'' for all private key generation algorithms and also the length of generated key is kept with the same.

\newcommand{\tabincell}[2]{\begin{tabular}{@{}#1@{}}#2\end{tabular}} 
\begin{table*}[htbp]
	\centering   
	\caption{The P-value of each key under four metrics.}
	\begin{tabular}{ccccc}
		\toprule
		\diagbox{\tiny{Method}}{\tiny{Metrics}}&\tabincell{c}{Non-overlapping\\Template Matching}&\tabincell{c}{Binary Matrix\\Rank}&\tabincell{c}{Maurer's ``Universal\\Statistical''}&\tabincell{c}{Random Excursions\\Variant}\\
		\hline
		Chaotic System &0&0.0504&0.2098&{\bf 0.1316}  \\
		LCG &0&0.9799&0.1966&0 \\
		Mersenne Twister &0&0.8106&0.1964&0  \\
		Rivest Cipher 4 &0&0.6828&0.1960&0.0857\\
		RSA &{\bf 0.9998}&5.22e-88&0&0\\
		DeepKeyGen(Ours) &0.5671&{\bf 0.9999}&{\bf 0.9931}&0.0517\\
		\bottomrule
	\end{tabular}\vspace{0cm}
	\label{tab-comp}
\end{table*}

The experimental results are shown in Table \ref{tab-comp}. It can be found that only the proposed method can always achieve a good performance on all evaluation metrics, where most algorithms just show their randomness under two metrics. Moreover, the P-value of the proposed method achieves the best performance than other private key generation algorithms in the term of binary matrix rank and maurer's ``universal statistical''. It can be concluded that the private key generated by the proposed DeepKeyGen achieves a better performance in the term of randomness. 

Overall, the proposed DeepKeyGen has the ability to generate the private key with higher security by comparing with existing private key generation algorithms. 

On the other hand, the experiments about the generating time for these algorithms are implemented to evaluate the efficiency. As shown in Table \ref{tab-time}, it can be found that the generating time of proposed DeepKeyGen is neither the longest nor the shortest. It costs about 2 seconds to generate a private key. The reason behind this is that the proposed DeepKeyGen is with a larger key space than other generation algorithms. The larger key space represents a higher security but it costs more time to calculate. By both considering the effectiveness and efficiency, the proposed DeepKeyGen can be a better choice to generate the private key. 

\begin{figure}[H]
	\centerline{
		\includegraphics[width=\columnwidth]{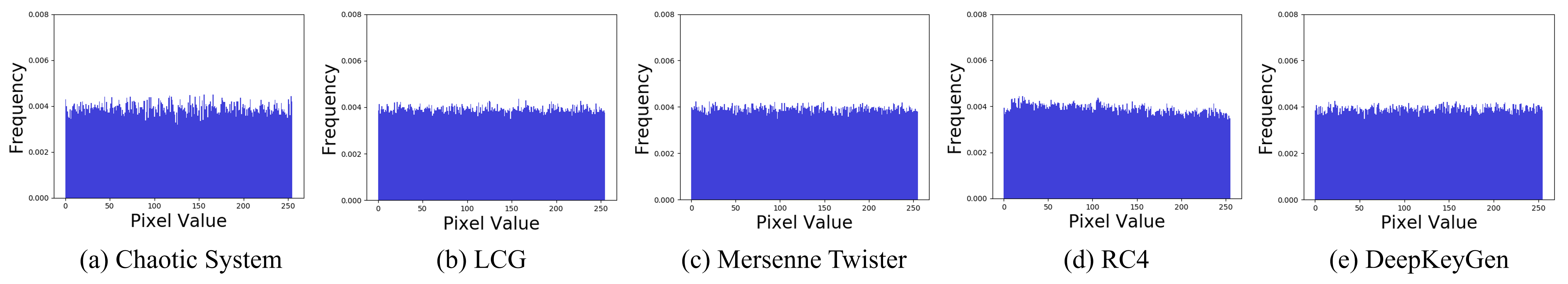}
	}
	\caption{Histogram of ciphertext images encrypted by different algorithms.}
	\label{fig-com}
\end{figure}

Furthermore, the histogram and information entropy are adopted to evaluate the quality of ciphertext images encrypted by different private keys. As seen in Table \ref{tab-com-entro}, the values of information entropy for different ciphertext images are almost the same and close to the 7.999. As shown in Fig.\ref{fig-com}, it also can be found that all encrypted images are with a uniform histogram distribution. It can eliminate the statistics of plaintext image. Based on the quality analysis of the ciphertext image, it can be said that private keys generated by these algorithms can be used to encrypt the plaintext image in a secure way. And it also can be proven that the proposed DeepKeyGen is an effective way to generate the private key like other private key generation algorithms. 

\begin{table}[htbp]
	\centering
	\fontsize{5}{7}\selectfont    
	\caption{Information entropy of private keys generated by different algorithms.}
	\begin{tabular}{ccccccc}
		\toprule
		Method&Chaotic System&LCG&Mersenne Twister&RC4&RSA&DeepKeyGen(Ours) \\
		\hline
		Entropy &7.9971&7.9991&7.9955&7.9990&6.0314&7.9870  \\
		\bottomrule
	\end{tabular}\vspace{0cm}
	\label{tab-com-key-entro}
\end{table}

\begin{table}[htbp]
	\centering
	\fontsize{5}{7}\selectfont    
	\caption{The key generating time of different algorithms.}
	\begin{tabular}{ccccccc}
		\toprule
		Method&Chaotic System&LCG&Mersenne Twister&RC4&RSA&DeepKeyGen(Ours) \\
		\hline
		TIME(s) &2.6178&0.5257&69.219&0.0649&0.4379&2.0714  \\
		\bottomrule
	\end{tabular}\vspace{0cm}
	\label{tab-time}
\end{table}

\begin{table}[htbp]
	\centering
	\fontsize{5}{7}\selectfont    
	\caption{Information entropy of ciphertext encrypted by different algorithms.}
	\begin{tabular}{ccccccc}
		\toprule
		Method&Chaotic System&LCG&Mersenne Twister&RC4&RSA&DeepKeyGen(Ours) \\
		\hline
		Entropy &7.9974&7.9991&7.9990&7.9990&7.2034&7.9986  \\
		\bottomrule
	\end{tabular}\vspace{0cm}
	\label{tab-com-entro}
\end{table}

\section{Performance Evaluation}

\subsection{Hyper-parameters}

In order to evaluate different hyper-parameters for training the network, experiments with different training epochs, learning rates and batch sizes are evaluated on the Montgomery County chest X-ray dataset. Moreover, this dataset is split into two parts, where 90\% dataset is adopted as as the training dataset and 10\% dataset is used as the validation dataset. In this study, the average value of information entropy for generated private keys is adopted as the evaluation metric to compare the different network hyper-parameters. The experimental results are shown in Table \ref{tab-hype}. 

It can be found that, when the learning rate is set with 0.02, the generated private key achieves a lower value for the information entropy. Furthermore, the private key cannot be generated in most training epochs when the batch size is set with 1. It means that the network training process cannot converge and the proposed generation network is unable to generate the private key. While the batch size is set with 6 or 10, the proposed network seems to be collapsed even if increasing the training epochs. And the information entropy of generated private key remains with the same value. As a whole, when the learning rate is set as 0.02, the proposed network is hard to achieve a good performance for generating the private key. Moreover, it can be found that if the learning rate is set with a smaller value, the network performance becomes better and better. To be more specific, the information entropy of generated private key can achieve can be improved under the learning rate of 0.002, but it is still hard to meet the security expectation for the private key, where the value should be close to 7.9. It also can be found that the network performance continues to increase with the decrement of the learning rate. When the learning rate becomes to 0.0002, most information entropy of generated private key is greater than the 7.9. And the best one (the information entropy is 9.9798) is achieved by setting with the 20000 training epochs and 1 for the batch size. Furthermore, under the learning rate of 0.0002, the network performance with batch size of 1 is always better than the result obtained from the batch size of 6 or 10. It can be said that, a smaller batch size holds the potential to facilitate the training process of private key generation network. Over, the best performance can be achieved when setting the learning rate with 0.0002, the training epoch with 20000 and the batch size with 1.

\renewcommand{\arraystretch}{1.2} 
\begin{table}[htbp]  
	\centering  
	\fontsize{7}{8}\selectfont  
	\caption{Study of Network Hyper-Parameters on Validation Dataset: the average value of information entropy for generated keys on different training epochs, learning rates and batch sizes.}  
	\label{tab-hype}  
	\setlength{\tabcolsep}{1mm}{
		\begin{tabular}{|c|c|c|c|c|c|c|c|c|c|}  
			\hline  
			\multirow{2}{*}{Epoch}&  
			\multicolumn{3}{c|}{LR=0.02}&\multicolumn{3}{c|}{LR=0.002}&\multicolumn{3}{c|}{LR=0.0002}\cr\cline{2-10}  
			&BS=1&BS=6&BS=10&BS=1&BS=6&BS=10&BS=1&BS=6&BS=10\cr\hline   
			10000&0.5222&3.7878&3.8766&5.1698&5.9714&6.1303&7.9380&7.7889&7.9004 \cr\hline  
			15000&NaN&3.7878&3.8766&4.6602&2.6587&5.7981&7.9405&7.7093&7.1990 \cr\hline  
			20000&NaN&3.7878&3.8766&5.4456&2.7983&6.1169&{\bf 7.9798}&7.9533&7.0686 \cr\hline  
			25000&NaN&3.7878&3.8766&5.1133&2.4782&5.6927&7.9490&7.1992&7.6847 \cr\hline  
			30000&NaN&3.7878&3.8766&7.9003&3.1811&6.6174&7.9555&6.7871&7.4846 \cr\hline  
		\end{tabular}  
	}
	\begin{tablenotes}   
		\scriptsize            
		\item[1]*LR represents the learning rate. BS represents the batch size. NaN represents not-a-number.
	\end{tablenotes} 
\end{table}

\subsection{Performance of Image Encryption and Decryption}

\begin{figure}[htbp]
	\centering
	\subfigure[]{
		\includegraphics[width=\columnwidth]{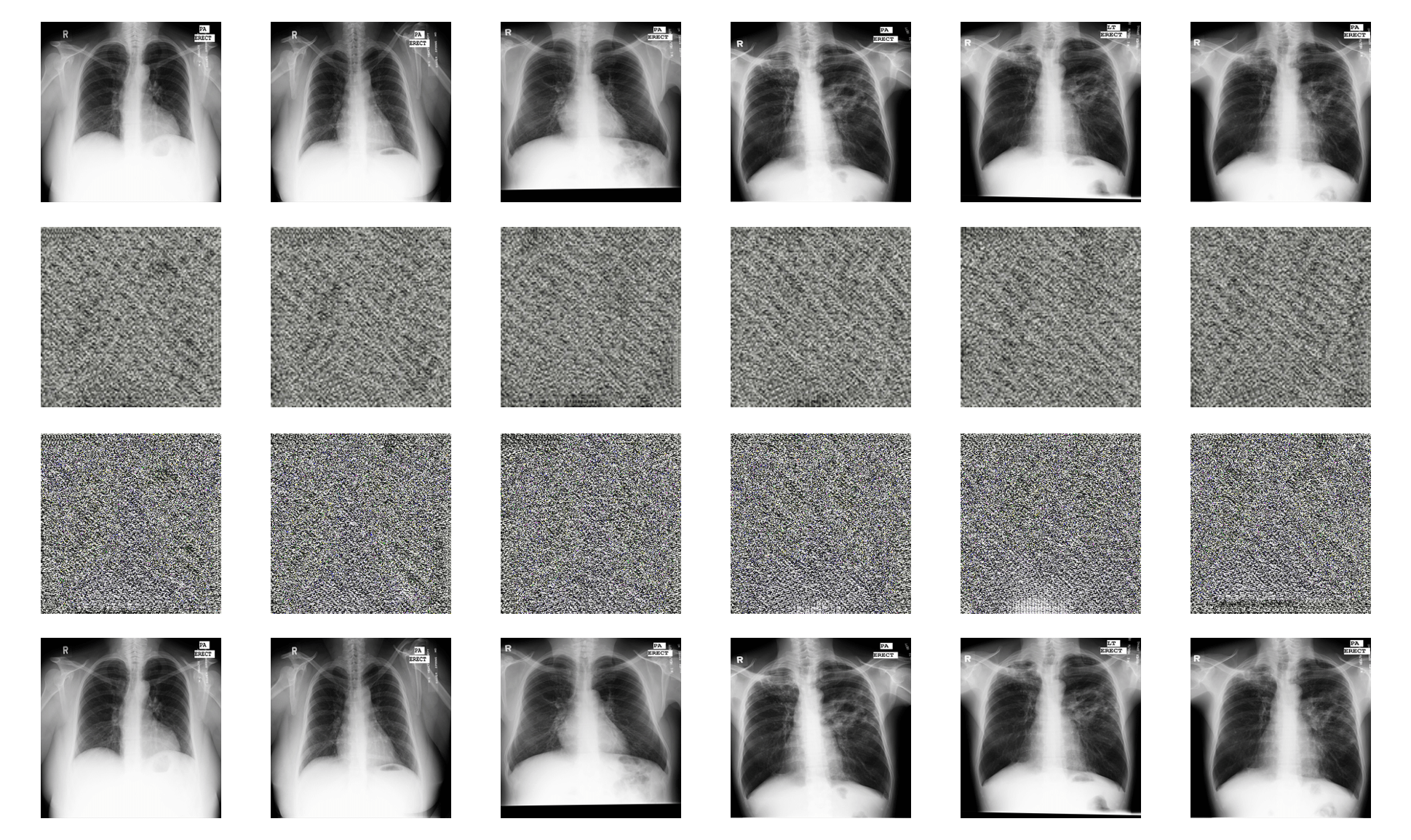} 
	}
	\subfigure[]{
		\includegraphics[width=\columnwidth]{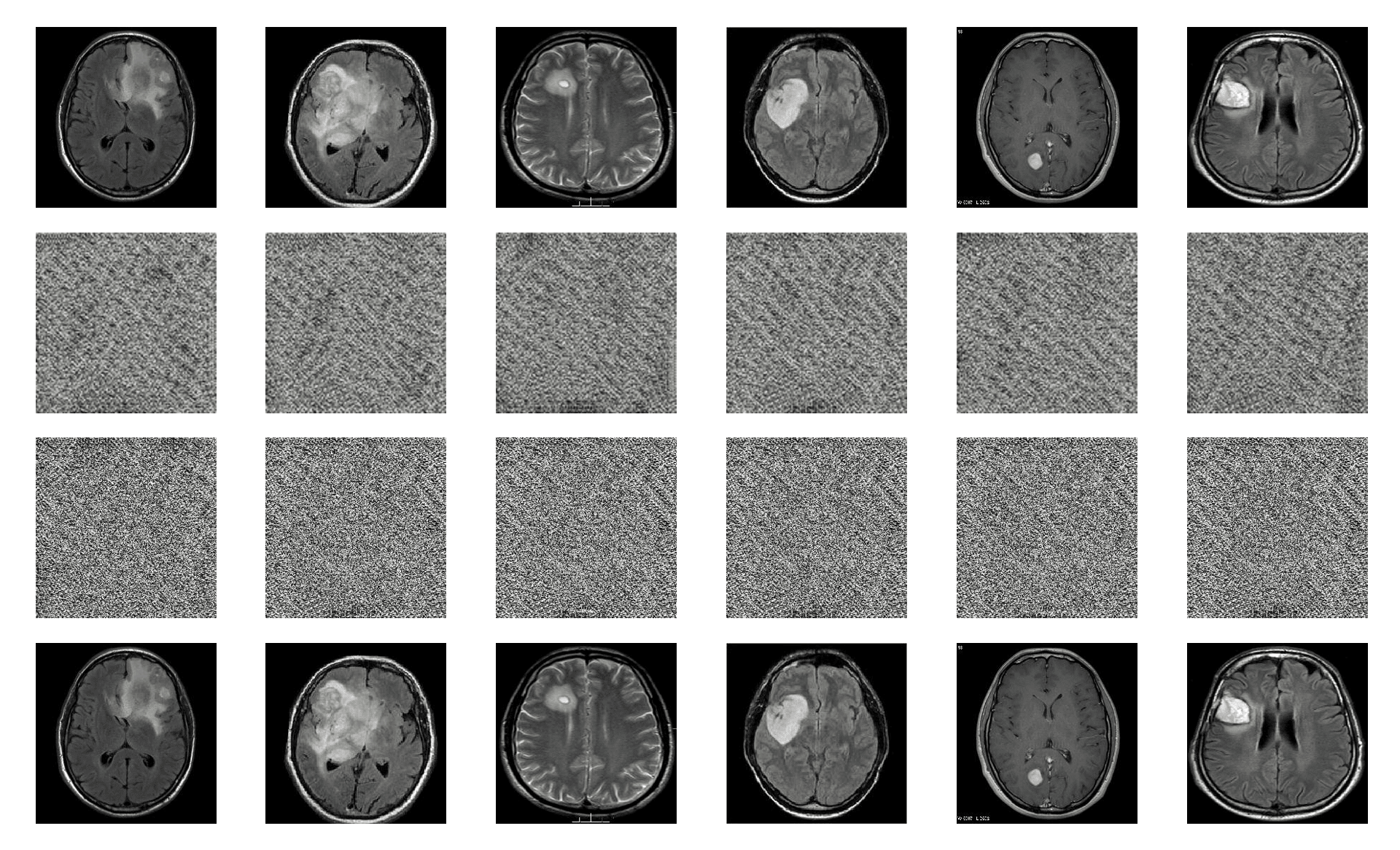} 
	}
	\subfigure[]{
		\includegraphics[width=\columnwidth]{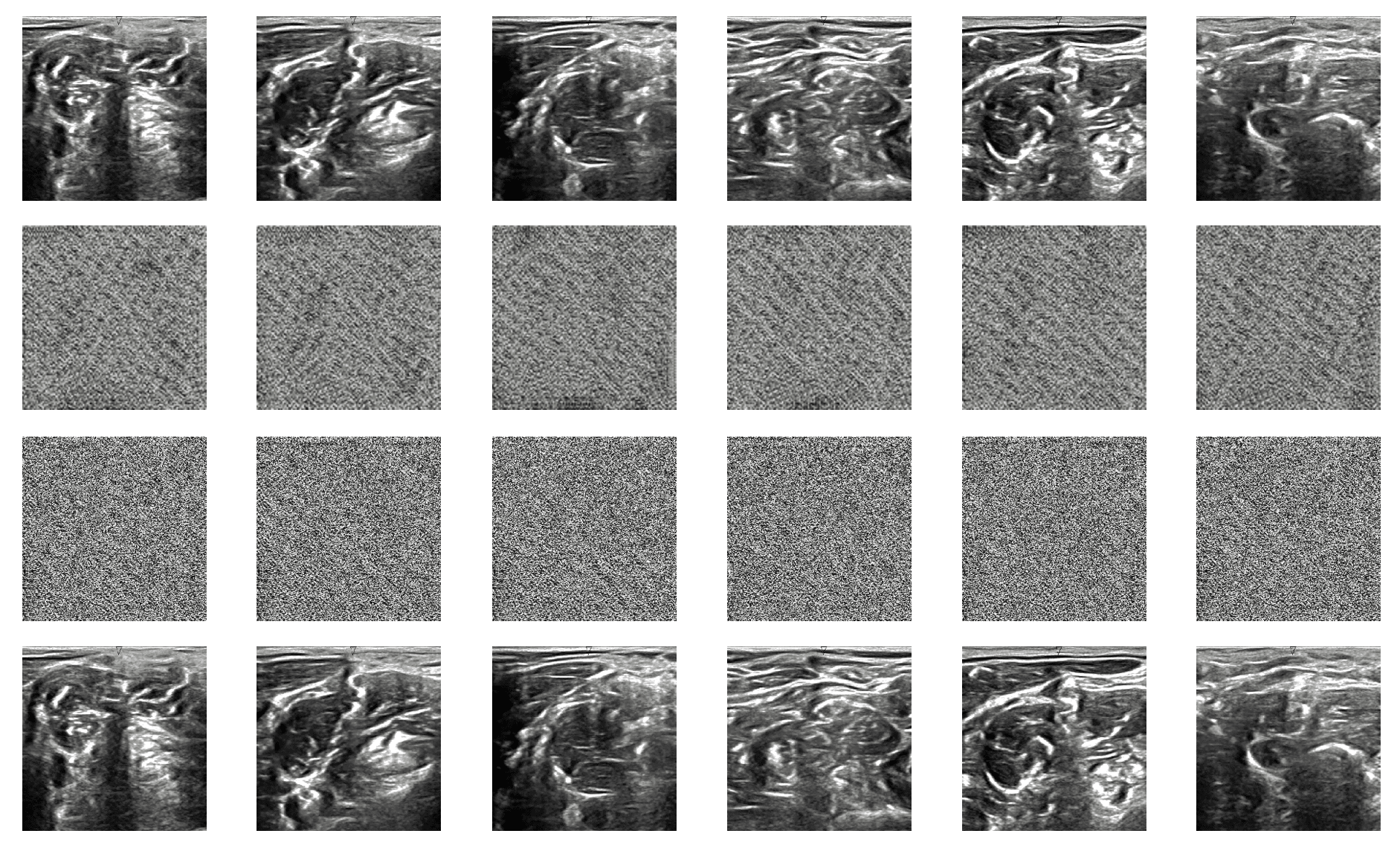} 
	}
	\caption{The qualitative analysis of the encryption and decryption performance are evaluated on (a) Montgomery County chest X-ray dataset, (b) Ultrasonic Brachial Plexus dataset and (c) BraTS18 dataset.}
	\label{fig14}
\end{figure}

The qualitative analysis of the encryption and decryption performance are evaluated on Montgomery County chest X-ray dataset, Ultrasonic Brachial Plexus dataset and BraTS18 dataset, and experimental results are presented in Fig.\ref{fig14} (a), (b) and (c) respectively. In these three figures, the first row represents the original medical images. The second row represents the private key generated by the proposed DeepKeyGen. The third row represents the ciphertext images encrypted by the generated keys and the fourth row represents the images decrypted by the private keys.

It can be found that the encrypted images are totally different from the original medical image and holds the ability to protect the patients' privacy in the medical image. Furthermore, the ciphertext image can be restored to the original image to realize the decryption process. According to the experimental result, it can be proven that the proposed DeepKeyGen is an effective key generation method to encrypt and decrypt the medical images with high security, which facilitate the process of protecting the private information of medical images. It also can be found that the proposed DeepKeyGen can be used to encrypt multi-modality medical images from different inspection equipment. Note that there is no correspondence between private key and the plaintext image. The generated private key can be used to encrypt any plaintext images by adopting different encryption/decryption algorithms.

\subsection{Impact on Transformation Domain}

\begin{figure}[H]
	\centering
	\subfigure[]{
		\includegraphics[width=0.10\textwidth]{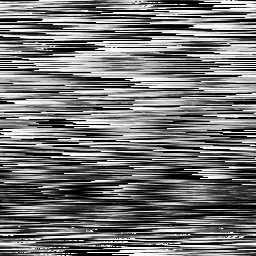} 
	}
	\subfigure[]{
		\includegraphics[width=0.10\textwidth]{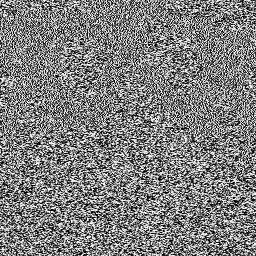} 
	}
	\subfigure[]{
		\includegraphics[width=0.10\textwidth]{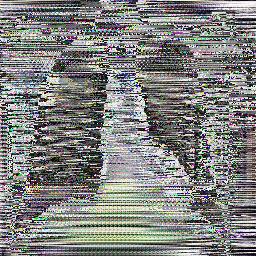} 
	}
	\subfigure[]{
		\includegraphics[width=0.10\textwidth]{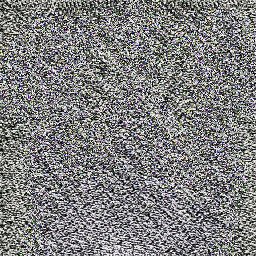} 
	}
	\caption{The encryption and decryption performance using different transformation domains. (a) Low-security transformation domain A. (b) High-security transformation domain B. (c) The encryption performance used the transformation domain A. (d) The encryption performance used the transformation domain B.}
	\label{fig17}
\end{figure}

The proposed DeepKeyGen is used to generate the private keys in an automatic way but under the guidance of the transformation domain. It means that the DeepKeyGen is designed to learn the expected ``style'' of the private key represented by the transformation domain. Therefore, the security of the generated keys mainly depends on the security performance in the transformation domain. In order to evaluate this dependency relationship, two different transformation domains A and B are adopted as a comparison. One transformation domain includes a set of images with the average information entropy of 7.1306 (low security), while the transformation domain is 7.9971 (high security). These two transformation domains are shown in Fig.\ref{fig17} (a) and Fig.\ref{fig17} (b) respectively.

For the transformation domain A with low security, it can be found that the average information entropy of the generated private keys is 7.8713 and the ciphertext images is 7.9865. While the value is 7.9798 for the private keys and 7.9986 for the ciphertext images when the transformation domain B is used. Moreover, the counter of the chest is exposed in the ciphertext image by employing the transformation domain A, which results in the leakage of medical information, but transformation domain B can protect the medical information well, as shown in Fig.\ref{fig17} (c) and Fig.\ref{fig17} (d) respectively. It indicates that the performance both on private keys and encrypted image from transformation domain A is worse than the result from transformation domain B. It proves the dependency relationship between the DeepKeyGen and the transformation domain. It also proves that the proposed DeepKeyGen has the ability to stably learn the mapping relationship from the source domain to the transformation domain and to guide the private key generation process to achieve the expected ``style'' of the private key presented in the transformation domain. It can be concluded that overall, the high security transformation domain brings high security to the privates keys and also a good encryption performance for the ciphertext image.

\section{Conclusion}
In this paper, we proposed a novel deep learning-based stream cipher generator, DeepKeyGen, which is designed to automatically generate the private key by directly learning the desirable ``style''. The proposed DeepKeyGen uses the images in the transformation domain as the desired ``style'' of the private key, and adopts the learning network to translate the initial image (``seed'') in the source domain onto the transformation domain, in order to generate the private key. The generated private key is then used to encrypt/decrypt medical image(s), by employing the XOR algorithm. DeepKeyGen was evaluated using the Montgomery County chest X-ray dataset, the Ultrasonic Brachial Plexus dataset, and the BraTS18 dataset. Extensive experimental results and security analysis on the private key and ciphertext image show that the stream cipher generated by the proposed DeepKeyGen has a large key space, pseudo-randomness, one-time pad, highly sensitive to change, and can resist different attacks. Compared with other key generation algorithms, DeepKeyGen achieves a high level of security. Moreover, using DeepKeyGen to encrypt multi-modality medical images achieves good performance. 

In future, our research direction will focus on how to adopt lightweight deep learning networks, such as MobileNet or Xception, to improve the efficiency of DeepKeyGen. We also plan to evaluate the security and performance of DeepKeyGen in different application domains to determine its generalizability. Furthermore, we will also make our protoype and relevant materials (e.g., codes) open source.


\ifCLASSOPTIONcaptionsoff
  \newpage
\fi

\end{document}